**Influence of Chloride Substitution on the Rotational Dynamics of Methylammonium in MAPbI$_{3-x}$Cl$_x$ Perovskites**

Götz Schuck[1]*, Frederike Lehmann[1,2], Jacques Ollivier[3], Hannu Mutka[3], and Susan Schorr[1,4]

[1] Helmholtz-Zentrum Berlin für Materialien und Energie, Hahn-Meitner-Platz 1, 14109 Berlin, Germany

[2] Universität Potsdam, Institut für Chemie, Karl- Liebknecht- Straße 24-25, 14476 Golm, Germany

[3] Institut Laue-Langevin, 71 Avenue des Martyrs, F-38000 Grenoble, France

[4] Institut für Geologische Wissenschaften, Freie Universität Berlin, Malteserstr. 74, 12249 Berlin, Germany

**Abstract**

The hybrid halide perovskites MAPbI$_3$, MAPbI$_{2.94}$Cl$_{0.06}$, and MAPbCl$_3$ (MA - methylammonium) have been investigated using inelastic and quasielastic neutron scattering (QENS) with the aim of elucidating the impact of chloride substitution on the rotational dynamics of MA. In this context, we discuss the influence of the inelastic neutron scattering caused by low-energy phonons on the QENS resulting from the MA rotational dynamics in MAPbI$_{3-x}$Cl$_x$. Through a comparative temperature-dependent QENS investigation with different energy resolutions, which allow a wide Fourier time window, we achieved a consistent description of the influence of chlorine substitution in MAPbI$_3$ on to the MA dynamics. Our results show that chlorine substitution in the low temperature orthorhombic phase leads to a weakening of the hydrogen bridge bonds since the characteristic relaxation times of C$_3$ rotation at 70 K in MAPbCl$_3$ (135 ps) and MAPbI$_{2.94}$Cl$_{0.06}$ (485 ps) are much shorter than in MAPbI$_3$ (1635 ps). For the orthorhombic phase, we obtained the activation energies from the temperature-dependent characteristic relaxation times $\tau_{C3}$ by Arrhenius fits indicating lower values of $E_a$ for $MAPbCl_3$ and $MAPbI_{2.94}Cl_{0.06}$ compared to $MAPbI_3$. We also performed QENS analyses at 190 K for all three samples. Here we observed that $MAPbCl_3$ shows slower MA rotational dynamics than $MAPbI_3$ in the disordered structure.

**Introduction**

Even if great hopes are placed on the use of methylammonium ($[CH_3NH_3]^+$, MA) lead halide perovskites as a light-collecting active layer in solar cells,[1] it is becoming increasingly clear that many basic properties of this group of photovoltaic materials are not yet sufficiently understood.[2,3] $ABX_3$ perovskites show huge potential for element substitutions on A-, B- and X-site which lead to a broad variety of physical properties.[4] Our main field of interest is chloride substituted MA lead triiodide in which the A-cation (12-fold coordination) is an organic unit (MA), the B-cation (octahedral coordination) = $Pb^{2+}$ and the X-anion = $I_{3-x}Cl_x$. Recent investigations of X-ray diffraction data were used to establish the solubility limits in $MAPbI_{3-x}Cl_x$ which are 3 mol-% chlorine in $MAPbI_3$ and 1 mol-% iodine in $MAPbCl_3$.[5,6] In general, $MAPbI_{3-x}Cl_x$ photovoltaic layers have at least three beneficial effects compared to $MAPbI_3$: 1) Increased stability, 2) superior charge carrier diffusion length (with identical band gap) and 3) remarkably high open circuit voltages. These impressive findings recently raised interest in the Cl-substituted perovskites.[7,8]

The aim of this study was to get an insight into the interrelationship of the static and dynamic structure of $MAPbI_{3-x}Cl_x$ studying the temperature dependent MA molecular dynamics by means of quasielastic neutron scattering (QENS) investigations in order to understand the influence of chloride on the rotational dynamics of the MA cation. On the basis of QENS experiments,[9-12] both the time constants and the movement model of the rotational dynamics taking place in the investigated $MAPbI_{3-x}Cl_x$ perovskite can be determined. With this, the dynamic behavior of the MA molecules can be described.[13]

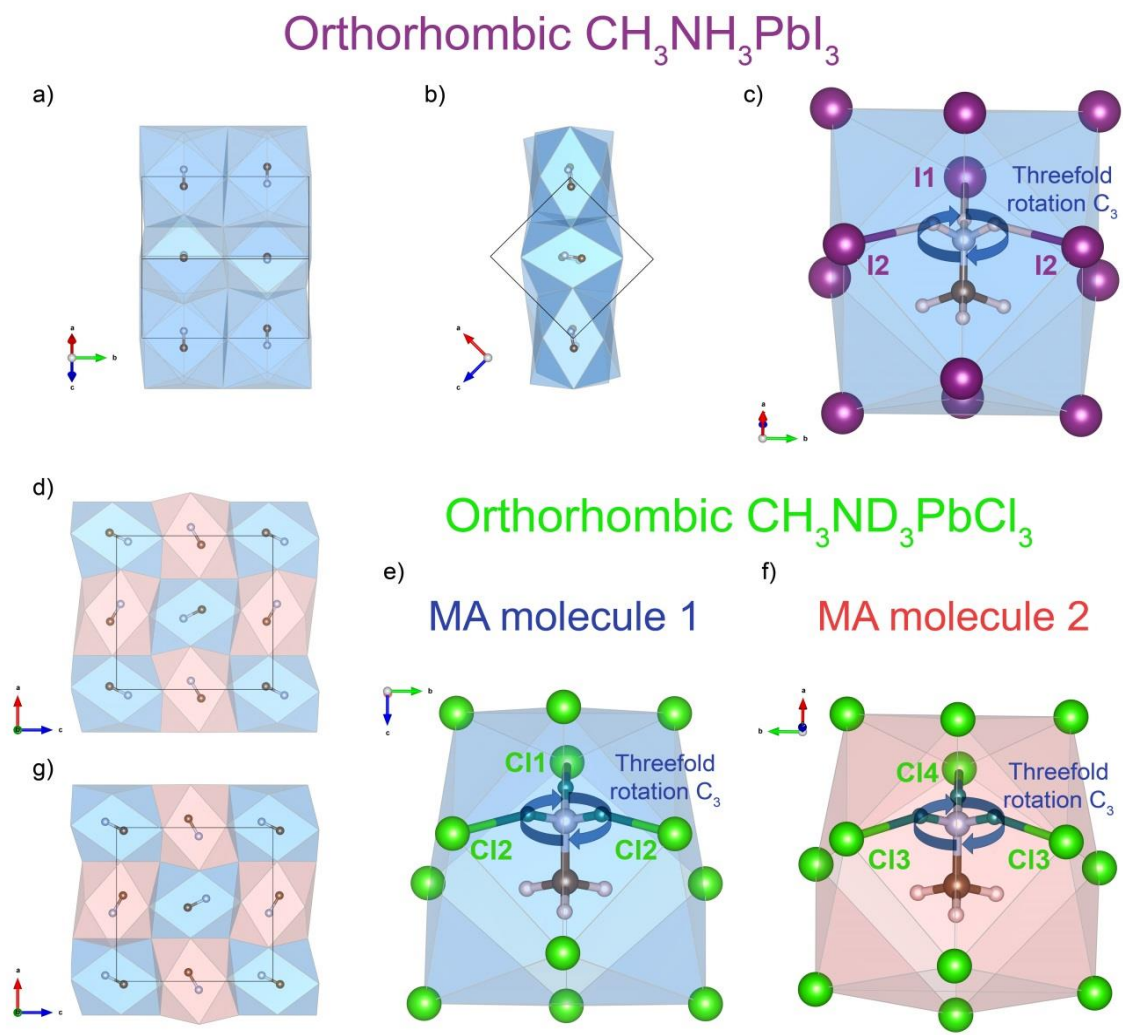

**FIGURE 1** Visualization of the orthorhombic crystal structures of $CH_3NH_3PbI_3$ (a, b, and c) at 100K [21] and $CH_3ND_3PbCl_3$ (d, e, f, and g) at 80 K [23] (note that for $MAPbCl_3$ only partially deuterated structural data is available) as well as the MA molecule $C_3$ jump rotation movement (c, e, and f). Due to the hydrogen (deuterium) bonds between $NH_3$ ($ND_3$) and the halide atoms in the orthorhombic crystal structures, the orientation of the C-N axes is fixed and leads to distortions of the crystal lattice. In a), b), d), and g) only the C-N atoms and their orientation are shown together with the respective 12-fold coordination polyhedra. Because of the doubled unit cell of $CH_3ND_3PbCl_3$ (compared to the iodide) two different MA molecules can be identified: MA molecule 1 (e; blue color) and MA molecule 2 (f; red color). In $MAPbI_3$, however, there is only one MA molecule (c) that occurs in four different orientations (a, b). Due to the double number of MA molecules in $CH_3ND_3PbCl_3$ there are a total of 8 different MA orientations (d, and g). In $CH_3ND_3PbCl_3$ the short N-Cl bond lengths are different in the two MA molecules: MA 1 (Cl2···N1 = 3.273(7) Å and Cl1-N1 = 3.336(11) Å) and MA 2 (Cl3···N2 = 3.300(6) Å and Cl4-N2 = 3.346(10) Å). Also the polyhedral volumes are different, $V_{MA1}$ = 155.8 Å$^3$, $V_{MA2}$ = 149.4 Å$^3$. In $MAPbI_3$ we can identify the following short N-I bond lengths: I2-N = 3.6804 Å and I1-N = 3.6113 Å. The polyhedral volume in $MAPbI_3$ is $V_{MA}$ = 206.9 Å$^3$. The figures c, e, and f also show the $C_3$ three-fold jump rotation movement (blue arrow) of the MA molecules around the C-N axis, which is discussed in section 2.1. The colors refer to the following elements: purple - iodine, green - chloride, brown - carbon, light blue - nitrogen, light orange - hydrogen, dark green - deuterium. All structural models are visualized by VESTA.[47]

In MAPbX$_3$ the crystal structure with the highest symmetry is the cubic aristotype with the space group $Pm\bar{3}m$. Here, the MA molecule can rotate freely inside the PbX$_6$ octahedra-corner-linked host structure. With decreasing temperature the PbX$_6$ octahedra-corner-linked host structure distorts more and more, at the same time the MA molecules can no longer rotate freely and start to form hydrogen bonds to halide atoms.[14-18,48] The ordering of the MA molecule and the PbX$_6$ host structure symmetry reduction results in tetragonal (*I4/mcm*) [19] and orthorhombic structural modifications (hettotypes).[13,20-24] Although the symmetry reduction leads to the same orthorhombic (*Pnma*) space group symmetry in MAPbI$_3$ and MAPbCl$_3$, the two low-temperature modifications differ considerably (Fig. 1). In the orthorhombic crystal structure of MAPbCl$_3$, which in comparison has a doubled unit cell size, four layers with short hydrogen bonds and two different MA molecules were identified instead of two layers and one MA molecule in MAPbI$_3$.[25] The phase transformations cubic to tetragonal and tetragonal to orthorhombic occur at 330 K (cubic → tetragonal) and 161 K (tetragonal → orthorhombic) for MAPbI$_3$ and at 177 K and 171 K for MAPbCl$_3$.[25] The phase transformation temperatures of MAPbI$_{2.94}$Cl$_{0.06}$ (326 K cubic → tetragonal and 155 K tetragonal → orthorhombic) are slightly lower compared to MAPbI$_3$ (Fig. S20). A similar temperature reduction of the phase transition temperatures was observed for MAPbI$_{2.7}$Br$_{0.3}$.[26] The transition from a freely rotating MA molecule in the cubic MAPbX$_3$ crystal structure via a disordered MA molecule arrangement in the tetragonal modification to the completely ordered arrangement of the MA molecules in the two orthorhombic low temperature structures is accompanied by drastic changes in the MA molecular dynamics and MAPbX$_3$ lattice oscillations. These temperature dependent changes of MA dynamics have been studied using various experimental techniques such as FTIR,[25,27-29] Raman spectroscopy,[30-33] inelastic neutron scattering,[34-38] and QENS.[13,36,39,40]

A comparative analysis of the temperature dependent inelastic neutron scattering (INS) of MAPbI$_3$, MAPbI$_{2.94}$Cl$_{0.06}$ and MAPbCl$_3$ in the orthorhombic low temperature structure was performed here for the first time. We observed that the incorporation of a low concentration of chlorine atoms into MAPbI$_3$ causes only small changes of the low-energy phonons. However, we also found that the INS spectra of pure MAPbCl$_3$ differ greatly from MAPbI$_3$. The QENS spectra of MAPbI$_3$ presented here are in very good agreement with already published results, but, our results show that it is important to cover a large observation time range [41] by using multiple energy resolutions in the QENS measurements to obtain a comprehensive picture of the MA molecule dynamics as a function of temperature. Furthermore, we show here for the first time that when the MA molecule motions become faster, an overlap with the low-energy phonons of the PbX$_6$ octahedron can no longer be excluded. We confirm the MA molecule motion models proposed by Chen et. al. [13] for the orthorhombic low-temperature structure and the tetragonal structure of MAPbI$_3$ and, for the first time, apply the same models to the QENS data of MAPbI$_{2.94}$Cl$_{0.06}$ and MAPbCl$_3$. Since temperature-dependent relaxation times $\tau_{C3}$ are determined for all three samples, the corresponding activation energies are also presented here and the resulting differences regarding the hydrogen bonds are discussed.

**Experimental details**

MAPbI$_3$, MAPbCl$_3$, and MAPbI$_{2.94}$Cl$_{0.06}$ was synthesized using a previously reported reaction which produced polycrystalline samples.[25] Sample purity was proven by X-ray powder diffraction analysis whereby no impurities were observed. Neutron scattering experiments were conducted on the neutron spectrometers IN5 [42] and IN4 at the Institute Laue-Langevin (ILL, Grenoble, France).[43] The IN5 experiments where conducted using three different incident neutron wavelengths of 2.25 Å (energy resolution of full width at half maximum (FWHM) = 800 µeV at the elastic line of Vanadium), 4.8 Å (energy resolution of FWHM =

86 µeV at the elastic line of Vanadium) and 8 Å (energy resolution of FWHM = 20 µeV at the elastic line of Vanadium) to provide three different time resolutions. IN5 data were collected at 70 K (2.25 Å and 4.8 Å: MAPbI$_3$, MAPbCl$_3$, and MAPbI$_{2.94}$Cl$_{0.06}$; 8 Å: MAPbI$_3$), 100 K (4.8 Å: MAPbI$_3$; 8 Å: MAPbI$_3$), 130 K (2.25 Å and 4.8 Å: MAPbI$_3$, MAPbCl$_3$, and MAPbI$_{2.94}$Cl$_{0.06}$), 160 K (4.8 Å: MAPbI$_3$; 8 Å: MAPbI$_3$) and 190 K (2.25 Å and 4.8 Å: MAPbI$_3$, MAPbCl$_3$, and MAPbI$_{2.94}$Cl$_{0.06}$). The IN4 experiments were conducted using an incident neutron wavelengths of 1.8 Å (energy resolution of FWHM = 1.5 meV at the elastic line), a continuous temperature scan from 10 K to 194 K over 860 minutes was carried out (Fig. S5), and 86 data sets with 10 minutes measuring time each were measured (heating rate 4.7 K per minute). The powder sample was placed in aluminum foil in an aluminum sample can and attached to the cold head of a helium cryostat. All sample handling was done under Argon atmosphere (Fig. S1). QENS was normalized to the signal from a vanadium standard and the inherent instrument background was corrected by subtracting an empty sample can. The following sample thicknesses result from the sample quantities used: 6.13 g / 1.47 mm (MAPbI$_3$), 4.85 g / 1.17 mm (MAPbI$_{2.94}$Cl$_{0.06}$) and 6.77 g / 2.13 mm (MAPbCl$_3$). This results in transmission values of T(90°) MAPbI$_3$ = 0.704, T(90°) MAPbI$_{2.94}$Cl$_{0.06}$ = 0.757 and T(90°) MAPbCl$_3$ = 0.678. No correction for multiple scattering was performed. Data were then converted to S(Q,ω). The reduction and analysis of the QENS data was carried out using the Large Array Manipulation Program (LAMP) software, developed by the computing group at ILL.[44] All fitting procedures were performed using the STRfit tool within the LAMP program. The Levenberg-G.Marquardt option (MPFIT based upon Minpack-1) was used as fit engine in the STRfit tool.

RESULTS AND DISCUSSION

**1. Inelastic Neutron Scattering (INS).** The inelastic neutron scattering we measured consists of the quasielastic neutron scattering, caused by the rotational jump of the MA molecule, and by the low energy optical phonon modes (Fig. 2).

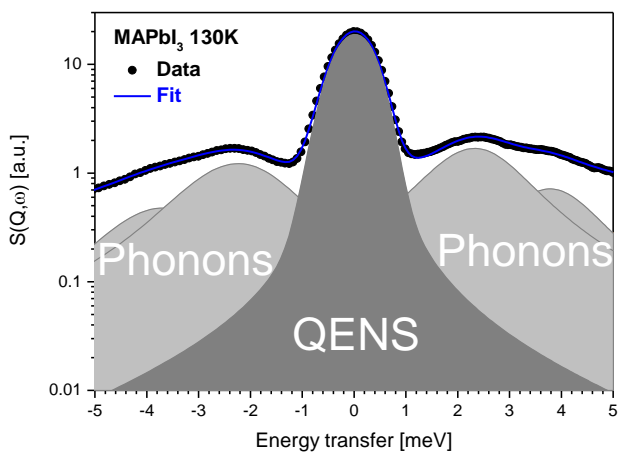

**Fig. 2** General features of QENS in MAPbI$_3$. An example of a resolution-convoluted fit of the quasielastic Lorentzian (QENS) and four Lorentzian functions in order to model the Phonons. Fit of the C$_3$ model to a low energy resolution spectrum measured at 130 K (IN5 2.25 Å; also shown as Figure S11a). Note the log scale for the intensities.

Looking at the inelastic neutron scattering spectra of MAPbI$_3$ measured at IN5, we see that there is a potential overlap between the low-energy phonons, in the energy range of 50 meV to 2 meV, and the quasielastic scattering caused by the rotational dynamics of the MA ions (Fig S2). The question is how the low-energy phonons behave as a function of temperature and whether they have a possible influence on the quasielastic scattering. To this end, we will first discuss the behavior of these low-energy phonons as a function of temperature for the three samples. Based on our factor group analysis [25] and based on literature data [28,30-38] we can state that the lattice vibrations of the lead atom and the two iodine atoms as well as the translation, libration and spinning [28] modes of the MA molecule can be observed in the energy range of 50 meV to 2meV. Furthermore, the internal MA vibration mode ν$_6$ (C-N torsion oscillation), which lies at an energy of around 37 meV, can also be observed in the inelastic neutron scattering data, but only with very low intensity (due to the low intensity, this mode was modeled separately). The phonons were modelled in the momentum integrated IN5 data in the range between 30 meV and 1.5 meV using Lorentzian functions (Fig. 3, S3 and S4). For the analysis, the momentum integrated IN5 data with 4.8 Å incident wavelength were used (temperatures: 70 K, 130 K and for MAPbI$_3$ also 100 K and 160 K). During the development of the fit model, it became apparent that it was necessary to specify a Lorentzian function at 7.5 meV with a fixed frequency (peak P6) for MAPbI$_3$ and MAPbI$_{2.94}$Cl$_{0.06}$ data. This assumption seems plausible since both Drużbicki et al. and Li et al. have observed phonon modes in the range of 6 to 9 meV.[35,36] Furthermore, the model also includes a Lorentzian function at 0 meV to map the influence of quasielastic scattering components. A complete summary of the determined lattice vibrations is given in Table S1 and S2.

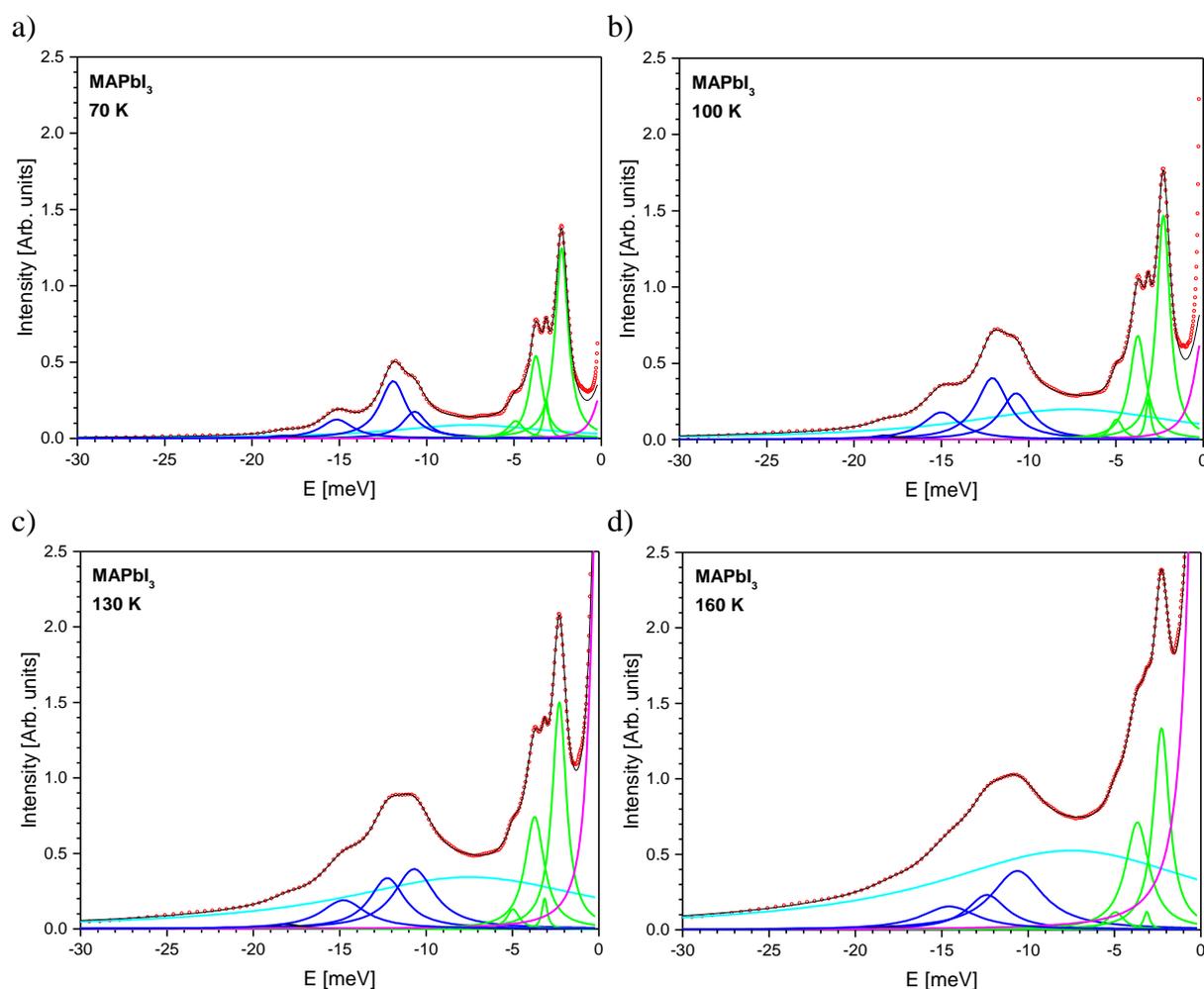

**Fig. 3** Momentum integrated INS spectra measured on IN5 (4.8 Å incident wavelength) of MAPbI$_3$ at 70 K (a), 100 K (b), 130 K (c) and 160 K (d). Lorentzian functions (decomposition of the vibrational modes according to Pérez-Osorio et. al. [28]): combination of MA translational and librational lattice vibrations without MA spinning (green), combination of MA translational and librational lattice vibrations with MA spinning (blue), MA lattice vibrations at fixed energy of -7.5 meV (cyan), and modelling the elastic and quasielastic neutron scattering at a fixed energy of 0 meV (magenta). Energy range used for the modelling: -30 to -1.5 meV. No background function was used here.

The observed phonon mode frequencies and half width at half maximum (HWHM) in the orthorhombic phase hardly differ for $MAPbI_3$ and $MAPbI_{2.94}Cl_{0.06}$, but large differences between $MAPbI_3$ and $MAPbCl_3$ can be observed. Two frequency ranges can be distinguished: The frequency range from 2.2 to 5.5 meV where the frequency and HWHM changes are small. And the frequency range from 10 meV to 20 meV. Here the changes are larger, the modes become considerably wider and the frequencies change partially in the opposite direction. Both the band assignment from, for example, Drużbicki et al. and the measurements from Swainson et al. on deuterated $MAPbBr_3$ indicate that, in the range from 2.2 to 5.5 meV, only oscillations of the lead halide are to be found whereas in the other frequency range from 7.5 to 30 meV the influence of the translation and libration modes of the MA molecule becomes stronger and stronger.[33-35] As has already been reported in published INS results,[34,35,38] a smearing effect caused by extremely strong broadening of the lattice oscillations at the transition from the orthorhombic to the tetragonal phase can be observed also in the IN5 data (Fig. S2). A softening of the oscillations of the lead halide as observed for $MAPbBr_3$ [34] can be excluded for $MAPbI_3$. The peak P1 at around 2.28 meV is not changing its frequency at least up to phase transition at 160 K. This result is also supported by other studies available.[38,45,46] The strong broadening effect in the tetragonal phase can not only be observed in $MAPbI_3$ but is also visible in the other two samples (Fig. S2). The analysis of momentum-integrated data of comparative $MAPbI_3$ measurements at IN4 with low energy resolution (Fig. S5) showed that this data can also be well described with the parameters obtained from the fit of IN5 4.8 Å data. In the fits of the IN4 data, the energy gain side is included in the analysis in addition to the energy loss (Fig. S6). If both the energy gain and the energy loss sides are taken into account, a description of the low-energy phonons by a constant flat background in a limited energy transfer range ±1 meV seems quite plausible (Fig. S7). For this reason we decided to consider the low-energy phonons for the higher (IN5 at 8 Å) and medium energy resolutions (IN5 at 4.8 Å) in the form of a constant flat background in the analysis of the quasielastic scattering (Fig. S9). For the 2.25 Å data, the inelastic background caused by the phonons is described with the help of additional Lorentzian functions; because both the energy loss and the energy gain side is accessible for the 2.25 Å data, and because for the QENS fits, a larger energy transfer range from ±7.5 meV to ±10 meV is used.

**2. Quasielastic neutron scattering (QENS).** In this QENS investigation of MAPbI$_3$, MAPbCl$_3$ and MAPbI$_{2.94}$Cl$_{0.06}$, the temperature and the energy resolution are varied in order to study the influence of chloride on the rotational dynamics of the MA molecule in these methylammonium lead halides.

**2.1 MAPbI$_3$ - orthorhombic phase and the C$_3$ jump model.** First we analyzed the temperature-dependent quasielastic neutron scattering (QENS) data of MAPbI$_3$ measured with 4.8 Å incident wavelength at temperatures below 161.5 K. In this temperature range the crystal structure of MAPbI$_3$ is orthorhombic (Fig. 1). We fixed both the elastic incoherent structure factor (EISF) and the quasielastic incoherent structure factor (QISF) to the suggested values based on the rotational jump models from Chen et. al. and Li.[13,36] For the orthorhombic phase, Chen and coworkers proposed that the quasielastic contribution can be described with only one quasielastic component QISF$_{C3}$ so that EISF = 1 - QISF$_{C3}$ (C$_3$ model).[13] The quasielastic contribution is described by a Lorentzian function L$_{C3}$ with QISF$_{C3}$ = $\frac{2}{3}$ [1 - $j_0$(Qr$\sqrt{3}$)],[12] where $j_0$ is the zeroth-order spherical Bessel function and r$\sqrt{3} = \overline{d_{H-H}}$ = 1.72 Å is the averaged distance between two hydrogen atoms in each of the methyl and ammonium groups.[13,36] The theoretical EISF and QISF behavior as a function of Q for the C$_3$ model is shown in Figure 5. From the HWHM $\Gamma_{C3}$ of the Lorentzian function L$_{C3}$, the correlation time $\tau_{C3}$ was obtained via the relation $\Gamma_{C3}$ [meV] $= \frac{3\hbar}{\tau_{C3}[ps]} = \frac{3*0.6583}{\tau_{C3}[ps]}$.[13]

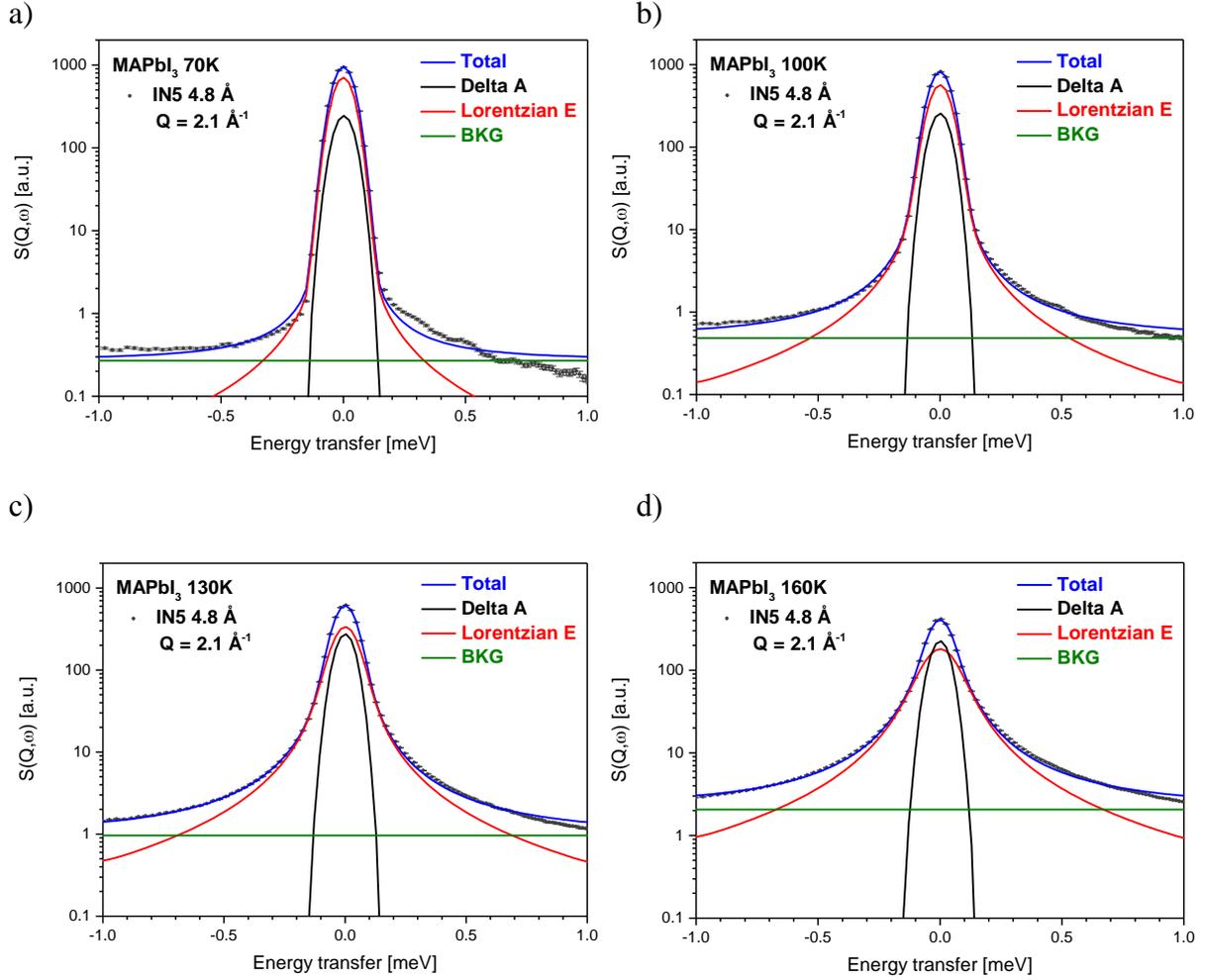

**Fig. 4** Jumping rotational dynamics of MA molecules in MAPbI$_3$. Fit of the C$_3$ model to a spectrum measured at a) 70 K, b) 100 K, c) 130 K, and d) 160 K (IN5 4.8 Å; elastic resolution FWHM = 86 μeV; Q = 2.1 Å$^{-1}$). The C$_3$ fit model includes: delta function (black line), Lorentzian function (red line), constant background (green line), and the sum (blue line) which fits the measured spectra (circles).

The energy transfer range from -1 meV to 1 meV was used to fit the Q dependent QENS data measured with 4.8 Å incident wavelength (Fig. 4). In a first step, the HWHM of the quasielastic component was determined simultaneously for 14 Q-values between 0.56 Å$^{-1}$ and 2.38 Å$^{-1}$ in a 2D fit. It was assumed here that the HWHM is Q-independent. Further fit parameters were the scaling factor, the peak shift of the elastic line and the constant background. The HWHM determined in the fit and the corresponding correlation times are listed in Table 1. In a second step, the HWHM was fixed and the ESIF determined experimentally (Fig. 5). Since we left the time resolution window at 70 K, a fit of the EISF for this temperature could not be performed. The deviations of the experimental EISF at small Q-values correspond to the behavior also observed by Li and coworkers.[36] However, since a single crystal (i.e. a thicker and more compacted sample) was used in the investigations of Li et. al.,[36] it cannot be ruled out that the EISF deviations observed there can be attributed to multiple scattering. Already, in the investigations of Chen et. al. (8 g powder was used here),[13] Li et. al.[36] did not make any corrections to the multiple scattering, so that it seems justified that a correction of the multiple scattering is omitted here for reasons of comparability. In the IN5 data with 4.8 Å incident wavelength Bragg reflexions occur at around Q = 2 Å$^{-1}$ (Fig. S8), so that the EISF is somewhat larger in this Q range. The background becomes larger with increasing Q values which can be explained by the increase of the inelastic scattering in the range of the low-energy phonons (Fig. S9). The temperature-dependent (70 K, 100 K and 160 K) IN5 QENS data which were measured with a higher energy resolution (8 Å) was evaluated in the same way (Fig. S10). The resulting HWHM and the corresponding correlation times are compiled in Table 1.

**Table 1** Jump model for the orthorhombic phase of MAPbI$_3$, MAPbCl$_3$ and MAPbI$_{2.94}$Cl$_{0.06}$. HWHM $\Gamma_{C3}$ of the Lorentzian function used to describe the quasielastic scattering for the corresponding temperatures and energy resolutions of the IN5 measurements. The correlation time $\tau_{C3}$ was obtained using $\Gamma_{C3}$ [meV] $= \frac{3\hbar}{\tau_{C3}[ps]} = \frac{3*0.6583}{\tau_{C3}[ps]}$.

| Sample | T [K] | HWHM $\Gamma_{C3}$ [μeV] | $\tau_{C3}$ [ps] | IN5 incident wavelength [Å] |
|---|---|---|---|---|
| **MAPbI$_3$** | 160 | 63.34 (3) | 31.2 | 4.8 |
| **MAPbI$_3$** | 130 | 26.498 (3) | 74.5 | 4.8 |
| **MAPbI$_3$** | 100 | 7.872 (5) | 251 | 4.8 |
| **MAPbI$_3$** | 70 | 1.208 (6) | 1635 | 4.8 |
| **MAPbI$_3$** | 160 | 63.00(6) | 31.3 | 8.0 |
| **MAPbI$_3$** | 100 | 8.132 (7) | 243 | 8.0 |
| **MAPbI$_3$** | 70 | 1.113 (2) | 1774 | 8.0 |
| **MAPbI$_3$** | 130 | 19 (9) | 104 | 2.25 |
| **MAPbI$_{2.94}$Cl$_{0.06}$** | 130 | 24.67 (1) | 80.1 | 4.8 |
| **MAPbI$_{2.94}$Cl$_{0.06}$** | 70 | 4.069 (9) | 485 | 4.8 |
| **MAPbI$_{2.94}$Cl$_{0.06}$** | 130 | 20 (8) | 98.7 | 2.25 |
| **MAPbCl$_3$** | 130 | 61.583 (3) | 32.1 | 4.8 |
| **MAPbCl$_3$** | 70 | 14.604 (6) | 135 | 4.8 |
| **MAPbCl$_3$** | 130 | 59.91 (8) | 33.0 | 2.25 |
| **MAPbCl$_3$** | 70 | 16.46 (8) | 120 | 2.25 |

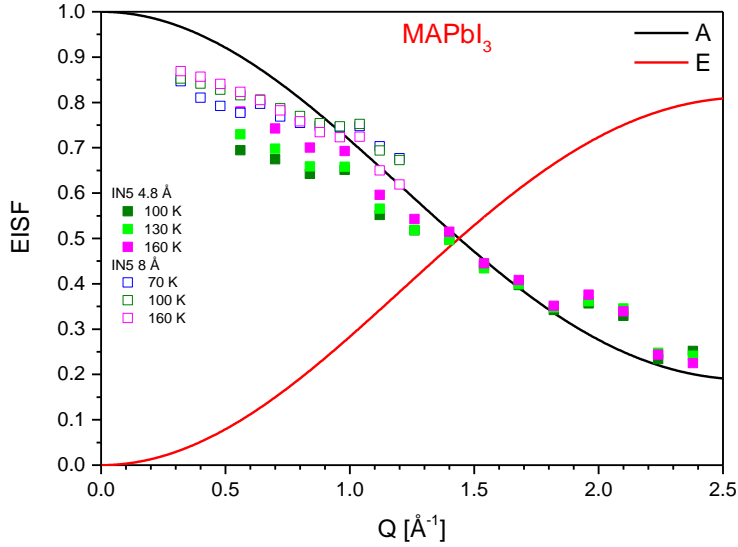

**Fig. 5** The experimental EISF of MAPbI$_3$ as a function of Q. At a temperature of 100 K (dark green), 130 K (green) and 160 K (magenta) for IN5 incident wavelength of 4.8 Å (filled symbols) and for 70 K (blue), 100 K (dark green) and 160 K (magenta) for IN5 incident wavelength of 8 Å (open symbols). The HWHM $\Gamma_{C3}$ compiled in Table 1 was used for the fit. The theoretical EISF (A, black line) and QISF (E, red line) behaviour is shown for comparison (the $C_3$ jump model described in section 2.1 was used to calculate the functions).

The fitted EISF values, together with the EISF values resulting from the low energy resolution measurements are shown in Fig. 5. In addition to the incident wavelength of 4.8 Å and 8 Å, QENS data measured at 2.25 Å were also analyzed for a temperature of 130 K (at 70 K, the quasielastic broadening is outside the time window of the 2.25 Å data, so that no QENS analysis could be performed). For the 130 K spectra for 14 Q-values between 1.0 Å$^{-1}$ and 4.25 Å$^{-1}$, fits were performed with the $C_3$ model in an energy transfer range of ±7.5 meV each (Fig. S11a). Now that the energy resolution is worse, the superposition of low-energetic phonons and QENS is much stronger. Therefore, it was necessary to include additional Lorentzian functions to describe the inelastic scattering caused by the phonons in the fit model instead of a constant background. The HWHM $\Gamma_{C3}$ resulting from the analysis for the 2.25 Å data corresponds within the error limits to the value determined for 4.8 Å (Table 1).

The experimental EISF values show that the $C_3$ model is very well suited for the description of QENS data even at high Q values (Fig. S11b). The small deviation of the experimental EISF values at high Q values, as it is the case with the experimental EISF values measured with 4.8 Å incident wavelength, shows the validity of the $C_3$ model for the interpretation of the QENS spectra of MAPbI$_3$. It seems that the deviations of the experimental EISF at high Q-values found by Li and coworkers were due to the fact that their QENS measurements used a too high energy resolution (FWHM 0.75 µeV) which is not able to capture the faster motion components of the MA molecule.[40] This can be seen particularly well in inelastic fixed window scans (IFWS) measurements where part of the motion components leave the measurable range at temperatures above 100 K (Fig. S7 in Li et. al. 2018).[40]

**2.2 MAPbI$_{2.94}$Cl$_{0.06}$ - orthorhombic phase.** The temperature-dependent IN5 data of MAPbI$_{2.94}$Cl$_{0.06}$ in the orthorhombic phase were analyzed analogous to the MAPbI$_3$ data. This means that first the HWHM $\Gamma_{C3}$ was determined in a 2D fit with fixed theoretical EISF and QISF, and then the experimental EISF was determined, whereby the HWHM $\Gamma_{C3}$ determined in the first step was given for all Q values. Even though only two temperatures (70 K and 130 K) were measured in the orthorhombic phase of MAPbI$_{2.94}$Cl$_{0.06}$ the 70 K data show remarkable differences to MAPbI$_3$. While the MAPbI$_{2.94}$Cl$_{0.06}$ QENS data at 130 K show almost the same half widths as the MAPbI$_3$ data (Tab. 1 and Fig. S12b), the QENS data at 70 K show quasielastic broadening that is almost four times wider than for the MAPbI$_3$ 70 K data (Fig. 6 and Fig. S12a). The experimental EISF values for MAPbI$_{2.94}$Cl$_{0.06}$ (Fig. S13) show a very similar behavior for both temperatures as the experimental EISF values determined for MAPbI$_3$ (Fig. 5). As with MAPbI$_3$, MAPbI$_{2.94}$Cl$_{0.06}$ also shows good agreement between the 2.25 Å and 4.8 Å data, the half widths $\Gamma_{C3}$ within the error are equal, and the experimental EISF values show good agreement with the $C_3$ model (Fig. S14 and Tab. 1).

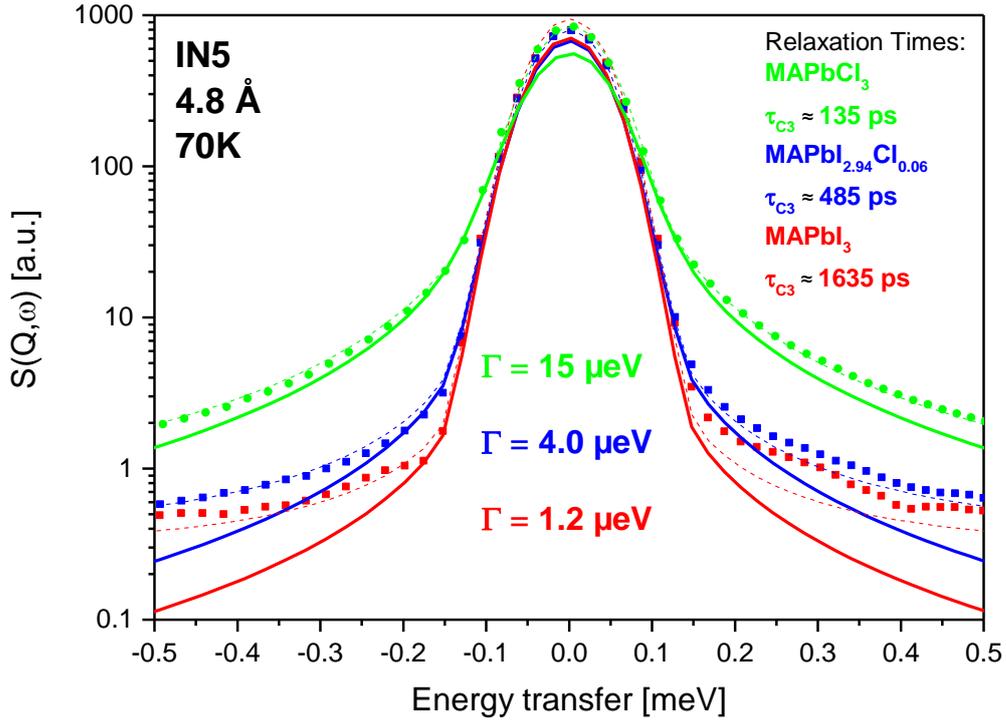

**Fig. 6** Jumping rotational dynamics of MA molecules in MAPbI$_3$ (red color), MAPbI$_{2.94}$Cl$_{0.06}$ (blue color), and MAPbCl$_3$ (green color). Fit of the C$_3$ model to spectra measured at 70 K (IN5 4.8 Å; elastic resolution FWHM = 86 μeV; Q = 2.1 Å$^{-1}$). Lorentzian function (solid line), and the sum (including the delta function and background; dashed line) which fits the measured spectra (squares).

Like for MAPbI$_3$ we found that at a temperature of 70 K, the quasielastic broadening is outside the time window of the 2.25 Å data, so that no QENS analysis could be performed for this temperature.

**2.3 MAPbCl$_3$ - orthorhombic phase.** The temperature-dependent IN5 data of MAPbCl$_3$ in the orthorhombic phase were also analyzed with the C$_3$ jumping rotational dynamics model analogous to MAPbI$_3$ and MAPbI$_{2.94}$Cl$_{0.06}$ (first the HWHM Γ$_{C3}$ was determined, with fixed theoretical EISF, and then the experimental EISF was determined). The quasielastic broadening is considerably wider for both 70 K and 130 K than for MAPbI$_3$ and MAPbI$_{2.94}$Cl$_{0.06}$. This is particularly visible in the 4.8 Å data which were measured at 70 K (Fig. 6). The more than 12 times wider broadening of MAPbCl$_3$ in comparison to MAPbI$_3$ can be clearly observed here. Due to the broader quasielastic component, both data sets (70K and 130 K) measured at 4.8 Å (Fig. S15) and both temperatures measured at 2.25 Å (Fig. S16) could be evaluated. For the 4.8 Å data a constant background was used for the description of the inelastic neutron scattering caused by the low-energy phonons, whereas for the 2.25 Å data additional Lorentzian functions were used. Even though the HWHM determined for the two energy resolutions are quite similar for the respective temperatures (Table 1), it is noticeable that with $MAPbCl_3$ the experimental ESIF for both temperatures and energy resolutions (Fig. S15c and S16c) shows greater deviations from the theoretical $C_3$ than with the other two investigated samples. The deviation in the QENS Fit at the 4.8 Å data measured at 130 K (Fig S15b) is also reflected in the larger deviations in the experimental EISF (Fig S15c). For the measurements at 4.8 Å, deviations in the opposite direction are visible (Fig S15c): On the one hand, the experimental EISF values are too large for the 130 K data at high Q values, and on the other hand, they are too small for the 70 K data at low Q values. However, since the 2.25 Å data at 130 K shows only small deviations from the $C_3$ model (Fig. S16c), an additional quasielastic component, as suggested by the experimental EISF deviations for the 4.8 Å data at 130 K, cannot be comprehended here. The decisive factor for the comparative QENS investigations, and thus for the validity of the $C_3$ model for the $MAPbCl_3$ 130 K data, is that the half-value widths for both energy resolutions at 130 K are sufficiently similar (Table 1). Opposite deviations of the experimental EISF values can also be observed for the 70 K data of $MAPbCl_3$. For the 70 K data measured with 4.8 Å, the experimental EISF values are too low, which could point to reduced mobility of the MA molecules, but for the data measured with 2.25 Å the experimental EISF values are too high. All in all, it is justified to assume an approximate validity of the C3 model for $MAPbCl_3$ in the orthorhombic phase and thus to open up the possibility of comparing the characteristic relaxation times of the three investigated samples. Due to the observed deviations of the experimental EISF values from the C3 model, it would be advisable to refine the two energy resolutions simultaneously (joint refinement), something which is unfortunately not yet possible with the used evaluation software.

### 2.3 The $C_3 \otimes C_4$ jump model.

As the only temperature above the orthorhombic structure, IN5 measurements at 190 K were performed for the two energy resolutions (4.8 Å and 2.25 Å incident neutron wavelengths) on all three samples. The $C_3 \otimes C_4$ model of Chen et. al. was used as the jump model for describing the quasielastic broadening at 190 K.[13] This model is based on the assumption that not only a three-fold jump component ($C_3$ jump model to describe the rotation along the C-N bond) but also a four-fold jump component $C_4$ (rotation perpendicular to the C-N bond) must be considered. The $C_3 \otimes C_4$ model includes one elastic component A$\otimes$A and five quasielastic components, convolution products ($\otimes$) of the $C_3$ and $C_4$ models components: A$\otimes$E, B$\otimes$A, B$\otimes$E, E$\otimes$A, and E$\otimes$E (corresponding to five Lorentzian functions). The behavior of these elastic and quasielastic incoherent structural factors $A_\gamma(Q)$ as a function of Q is expressed by a combination of a series of zero order spherical Bessel functions $j_i = j_0(Qr_i)$ where $r_i$ are the jump distances to be considered.[13] In the $C_3 \otimes C_4$ model it is assumed that the H-H distances for $CH_3$ and the H-H distances $NH_3$ are the same and correspond to an average value of 1.72 Å. The Q dependence of the elastic and quasielastic incoherent structural factors is shown in figure S17. The following relationships apply between the HWHM and the relaxation times:

$\Gamma_{A\otimes E} = \frac{3}{\tau_{C3}}$, $\Gamma_{B\otimes A} = \frac{4}{\tau_{C4}}$ , $\Gamma_{B\otimes E} = \frac{4}{\tau_{C4}} + \frac{3}{\tau_{C3}}$, $\Gamma_{E\otimes A} = \frac{2}{\tau_{C4}} + \frac{2}{\tau_{C2}}$, $\Gamma_{E\otimes E} = \frac{2}{\tau_{C4}} + \frac{2}{\tau_{C2}} + \frac{3}{\tau_{C3}}$ .

According to Chen et. al. only the two HWHM $\Gamma_{C3}$ and $\Gamma_{C4}$ are given so that $\Gamma_{C2}$ is obviously very small.[13] We can therefore establish the following relation: $\Gamma_{A\otimes E} = \Gamma_{C3}$, $\Gamma_{B\otimes A} = \Gamma_{C4}$, $\Gamma_{B\otimes E} = \Gamma_{C3} + \Gamma_{C4}$, $\Gamma_{E\otimes A} = 0.5\ \Gamma_{C4}$, and $\Gamma_{E\otimes E} = 0.5\ \Gamma_{C4} + \Gamma_{C3}$. Since $\Gamma_{C3}$ corresponds to a very fast movement with relaxation times of 1 to 2 ps and $\Gamma_{C4}$ corresponds to a much slower movement with times in the range of 100 ps, we have three broad Lorentz functions ($\Gamma_{A\otimes E}, \Gamma_{B\otimes E}, \Gamma_{E\otimes E}$) and two narrow ones ($\Gamma_{B\otimes A}$, $\Gamma_{E\otimes A}$) as we can see in a fit of the $C_3\otimes C_4$ model of 4.8 Å data of MAPbI$_3$ at 190 K (Fig. 7). A constant background was used in this fit and in the corresponding fits of the MAPbI$_{2.94}$Cl$_{0.06}$ (Fig. S18a) and MAPbCl$_3$ (Fig. S18b) spectra measured at 190 K. In comparison to the fits on the orthorhombic spectra, the energy transfer range was extended to -4 meV to 1.5 meV in the tetragonal (for MAPbCl$_3$ cubic) phase to describe the broader quasielastic components in the 4.8 Å data. The HWHM and relaxation times resulting from the $C_3\otimes C_4$ fits are listed in table 2. Here it can be seen that the relaxation times $\tau_{C3}$ for MAPbI$_3$ (1.4 ps) and MAPbI$_{2.94}$Cl$_{0.06}$ (1.35 ps) are similar whereas for MAPbCl$_3$ a longer relaxation time of 4.65 ps was found. This is also visible in the direct comparison of the spectra of MAPbI$_{2.94}$Cl$_{0.06}$ and MAPbCl$_3$ fitted with the $C_3\otimes C_4$ model since the wide components in MAPbI$_{2.94}$Cl$_{0.06}$ are more than three times as wide as in MAPbCl$_3$ (Fig. S18). In addition, half-widths could be determined from the fits of spectra measured with 2.25 Å (Table 2 and Fig. S19). As in the fits of the 4.8 Å data, it is evident that at a temperature of 190 K the MA $C_3$ movement component in MAPbCl$_3$ rotates slower than in MAPbI$_3$ and MAPbI$_{2.94}$Cl$_{0.06}$. For the $C_4$ component, the result is not so clear since a separation of the two motion components $C_3$ and $C_4$ is no longer obvious at lower energy resolution. The differences between the two energy resolutions can might also be explained by the fact that a larger energy transfer range (±7.5 meV) was used for the 2.25 Å data. Another difference is that, as with the other 2.25 Å data, Lorentzian functions were utilized to describe the low-energy phonons (six Lorentzian functions for MAPbI$_3$ and MAPbI$_{2.94}$Cl$_{0.06}$ and two Lorentzian functions for MAPbCl$_3$). Again, it is expected that simultaneous analysis of the 4.8 Å and 2.25 Å spectra would provide a more uniform result.

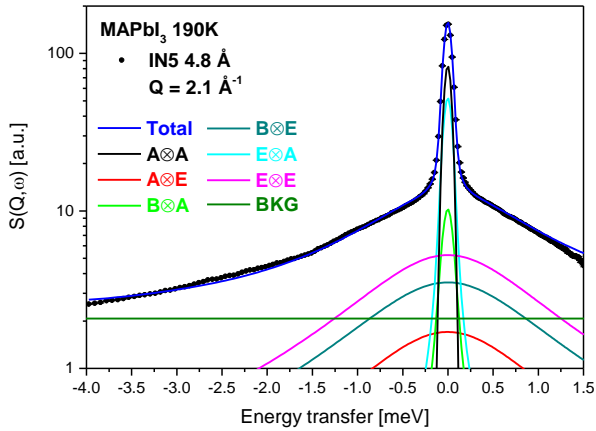

**Fig. 7** Jumping rotational dynamics of MA molecules in MAPbI$_3$. Fit of the C$_3 \otimes$C$_4$ model to a spectrum measured at 190 K (IN5 4.8 Å; elastic resolution FWHM = 86 μeV at Q = 2.1 Å$^{-1}$). The C$_3 \otimes$C$_4$ fit model includes: delta function (black line), five Lorentzian functions, constant background (dark green line), and the sum (blue line) which fits the measured spectra (circles).

**Table 2** Jump model for the tetragonal phase of MAPbI$_3$ and MAPbI$_{2.94}$Cl$_{0.06}$; and for the cubic phase of MAPbCl$_3$. HWHM $\Gamma_{C3}$ and HWHM $\Gamma_{C4}$ of the Lorentzian functions used to describe the quasielastic scattering for the corresponding temperature and energy resolutions of the IN5 measurements. The correlation time $\tau_{C3}$ was obtained using $\Gamma_{C3}$ [meV] $= \frac{3\hbar}{\tau_{C3}[ps]} = \frac{3*0.6583}{\tau_{C3}[ps]}$. The correlation time $\tau_{C4}$ was obtained using $\Gamma_{C4}$ [meV] $= \frac{4\hbar}{\tau_{C4}[ps]} = \frac{4*0.6583}{\tau_{C4}[ps]}$.

| Sample | T [K] | HWHM $\Gamma_{C3}$ [μeV] | $\tau_{C3}$ [ps] | HWHM $\Gamma_{C4}$ [μeV] | $\tau_{C4}$ [ps] | IN5 incident wavelength [Å] |
|---|---|---|---|---|---|---|
| **MAPbI$_3$** | 190 | 1407 (8) | 1.40 | 30.86 (3) | 85.3 | 4.8 |
| **MAPbI$_3$** | 190 | 1635 (4) | 1.21 | 9.9 (4) | 266 | 2.25 |
| **MAPbI$_{2.94}$Cl$_{0.06}$** | 190 | 1462(7) | 1.35 | 31.27 (3) | 84.2 | 4.8 |
| **MAPbI$_{2.94}$Cl$_{0.06}$** | 190 | 1566(4) | 1.26 | 7(1) | 376 | 2.25 |
| **MAPbCl$_3$** | 190 | 425(3) | 4.65 | 46(1) | 57.2 | 4.8 |
| **MAPbCl$_3$** | 190 | 1380(5) | 1.43 | 7.6 (7) | 346 | 2.25 |

**2.4 Activation energies for orthorhombic MAPbI$_3$, MAPbI$_{2.94}$Cl$_{0.06}$, and MAPbCl$_3$ and the influence of chlorine on hydrogen bridge bonding.**

In general, the strength of the hydrogen bonds can be described by various parameters.[25,49] From crystal structure data (see also Schuck et. al., Table S5)[25], for example, the bond angle of the hydrogen bond can be determined (largest angles N-H...X in the orthorhombic crystal structure: for MAPbCl$_3$ 158.8° [21] and for MAPbI$_3$ 174.1° [23]), which can then be used as a characteristic for the strength of the hydrogen bond. Furthermore, the H...X atom distances in comparison to mean atom distances from literature data can also be used as comparison parameters for hydrogen bonds (smallest H...X distances for the orthorhombic crystal structure: for MAPbCl$_3$ 2.292 Å [21] and for MAPbI$_3$ 2.613 Å [23] in comparison to literature mean values from Steiner [49] 2.24 Å for Cl$^-$ and 2.66 Å for I$^-$). However, the observations for the static crystal structures do not sufficiently describe the conditions in the hydrogen bond since the rotational jump molecule dynamics is very fast. Therefore, besides the already considered characteristic relaxation times of the C$_3$ jump rotation, the activation energies of the C$_3$ jump rotation are of great importance. For the characteristic relaxation times of C$_3$ jump rotation, the following relationship applies: the smaller the $\tau_{C3}$, the weaker the strength of the hydrogen bond, since the MA molecule can perform a faster jump rotation. Correspondingly, smaller activation energies indicate a weaker hydrogen bond. From the temperature-dependent behavior of the experimentally determined HWHM $\Gamma_{C3}$, the activation energies E$_A$ in the orthorhombic phase for the three halides can be determined according to the Arrhenius law (Fig. 8). In each case, all available energy resolutions were used to determine the activation energies, since it was observed that the HWHM determined in the different observation time ranges (energy resolutions) differ only slightly (Tab. 1). A direct comparison is possible since the same movement model C$_3$ was used for all three halides. First of all, the results indicate that the activation energy of E$_a$ = 41(2) meV determined for MAPbI$_3$ is in the range of the values reported by Chen et al. 48(3) meV and Li et al. 47.9(6) meV.[13,36] The activation energies of MAPbI$_{2.94}$Cl$_{0.06}$ E$_a$ = 22(2) meV and MAPbCl$_3$ E$_a$ = 17.8(8) meV are significantly lower compared to MAPbI$_3$. This shows the direct influence of chlorine on the dynamics of the MA jumping rotation. Even small amounts of chlorine in MAPbI$_3$ obviously weaken the hydrogen bond in the orthorhombic phase so that a faster MA dynamic is possible.

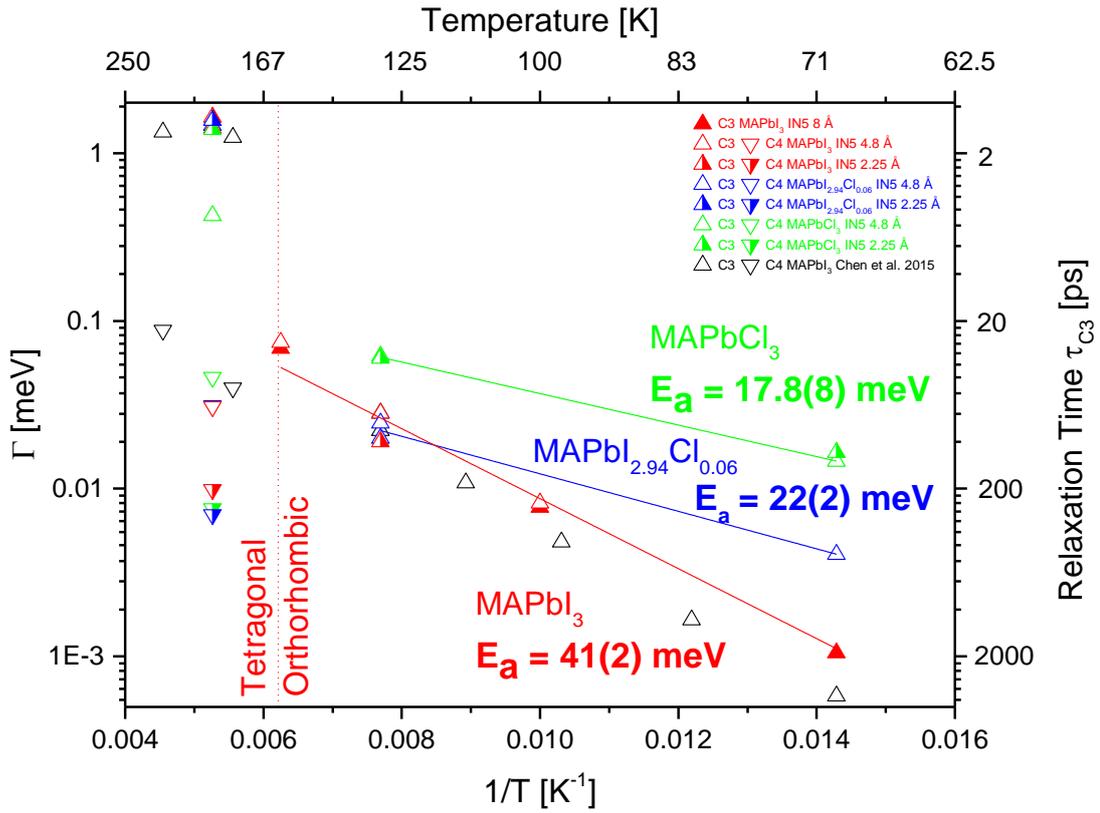

**Fig. 8** The temperature-dependence of HWHM $\Gamma_{C3}$ and $\Gamma_{C4}$ for MAPbI$_3$ (red symbols), MAPbI$_{2.94}$Cl$_{0.06}$ (blue symbols), and MAPbCl$_3$ (green symbols). In the orthorhombic phase, $\Gamma_{C3}$ is fitted to the Arrhenius relation $\Gamma_{C3} \propto (-\frac{E_a}{k_B T})$, where $k_B$ is the Boltzmann constant and $E_a$ the activation energy of the jumping rotational dynamics of MA molecule. For comparison, the values determined by Chen et. al. are also shown in the diagram (black symbols).[13]

The influence of chlorine in MAPbI$_3$ is particularly clear at low temperatures (Fig. 6). At 70 K, 2 % chlorine in MAPbI$_3$ ($\tau_{C3}$ = 485 ps) cause a more than three times faster jump rotation than in pure MAPbI$_3$ ($\tau_{C3}$ = 1635 ps). Corresponding with our investigations with FTIR,[25] we can state that the MA molecule in the iodide perovskite at low temperatures, in the ordered orthorhombic phase, is more strongly influenced by the hydrogen bridges than in the chloride perovskite or in chloride substituted MAPbI$_3$. Even if in the tetragonal phase above 160 K no activation energies could be obtained, conclusions can be drawn about the influence of chlorine on the MA dynamics. While in the orthorhombic low temperature phase of MAPbCl$_3$ and also of MAPbI$_{2.94}$Cl$_{0.06}$ chlorine causes the MA molecule to move faster than in MAPbI$_3$, our results indicate that in the tetragonal phase the $C_3$ component of the MA molecule jump rotation are faster in MAPbI$_3$ (1.40 ps) than in MAPbCl$_3$ (4.65 ps). This would mean that the influence of the hydrogen bonds in the disordered tetragonal phase of MAPbI$_3$ is weaker than in MAPbCl$_3$.

**Conclusions**

We studied temperature-dependent QENS spectra of a series of MA lead halide perovskites. Through the analysis of the inelastic neutron scattering data in the low-energy phonon range, comparative INS results could be obtained for the three investigated samples. The observed frequencies for MAPbI$_3$ match with the already published results.[36] For MAPbI$_{2.94}$Cl$_{0.06}$ and MAPbCl$_3$, INS results could be reported here for the first time. However, the INS data analysis also showed that the phonon scattering contributions must be taken into account in QENS analyses with lower energy resolutions. In the orthorhombic as well as in the tetragonal phase, a good agreement of the determined correlation time could be achieved for the spectra of MAPbI$_3$ in comparison to the already performed QENS investigations.[13,36] It is also remarkable that the good agreement in the use of up to three different energy resolutions could be achieved, especially in QENS data of the orthorhombic low temperature phase. In particular, the QENS data with low energy resolution (2.25 Å) contributed to showing that the $C_3$ model is valid in the orthorhombic phase of MAPbI$_3$ since here a good agreement of the experimental EISF values with the $C_3$ model could be achieved at large Q values. For MAPbI$_{2.94}$Cl$_{0.06}$ and MAPbCl$_3$, QENS measurements were carried out for the first time. Again, it was shown that the $C_3$ model in the orthorhombic phase and the $C_3 \otimes C_4$ model in the tetragonal phase (cubic phase for MAPbCl$_3$) describes the experimental data reasonably well. Nevertheless, we also see that especially with the measurements at 190 K larger differences occur between the measurements with moderate and broader energy resolution because the scattering components caused by phonons are much wider than at low temperatures and therefore more difficult to describe. However, as the quasielastic components become broaden at the same time, a simultaneous refinement of QENS data with different energy resolutions, with simultaneous better theoretical knowledge of the true intrinsic phonon half-widths, would allow much more consistent results.

ASSOCIATED CONTENT

The following are available free of charge.

**Supporting Information.** Table S1: Frequencies of peaks in INS spectra, Table S2: HWHM of peaks in INS spectra, Fig. S1: sample handling, Fig. S2: INS spectra $MAPbI_{2.94}Cl_{0.06}$ and $MAPbCl_3$, Fig. S3: INS spectra $MAPbI_{2.94}Cl_{0.06}$, Fig. S4: INS spectra $MAPbCl_3$, Fig. S5: INS spectra $MAPbI_3$, Fig. S6: INS spectra $MAPbI_3$, Fig. S7: Results of a straight-line fit, Fig S8: $MAPbI_3$ neutron powder diffraction simulation, Fig. S9 Background in QENS Fits, Fig. S10 QENS (IN5 8 Å) fit $MAPbI_3$ ($C_3$ model at 70 K, 100 K, 130 K, and 160 K), Fig. S11 QENS (IN5 2.25 Å) fit and EISF of $MAPbI_3$ ($C_3$ model at 130 K), Fig. S12 QENS (IN5 4.8 Å) fit $MAPbI_{2.94}Cl_{0.06}$ ($C_3$ model at 70K and 130 K), Fig. S13 EISF (IN5 4.8 Å) of $MAPbI_{2.94}Cl_{0.06}$ ($C_3$ model at 70K and 130 K), Fig. S14 QENS (IN5 2.25 Å) fit and EISF of $MAPbI_{2.94}Cl_{0.06}$ ($C_3$ model at 130 K), Fig. S15 QENS (IN5 4.8 Å) fit and EISF of $MAPbCl_3$ ($C_3$ model at 70 K and 130 K), Fig. S16 QENS (IN5 2.25 Å) fit and EISF of $MAPbCl_3$ ($C_3$ model at 70 K and 130 K), Fig. S17 Q dependence of the elastic and quasielastic incoherent structure factors for the $C_3 \otimes C_4$ model, Fig. S18 QENS (IN5 4.8 Å) fit $MAPbCl_3$ and $MAPbI_{2.94}Cl_{0.06}$ ($C_3 \otimes C_4$ model at 190 K), Fig. S19 QENS (IN5 2.25 Å) fit $MAPbI_3$, $MAPbCl_3$ and $MAPbI_{2.94}Cl_{0.06}$ ($C_3 \otimes C_4$ model at 190 K). (File type, PDF), Fig. S20 Phase diagram of $MAPbI_{3-x}Cl_x$ and Temperature-dependent synchrotron X-ray diffraction data of $MAPbI_{3-x}Cl_x$

AUTHOR INFORMATION

**Corresponding Author**

*Götz Schuck, e-mail: goetz.schuck@helmholtz-berlin.de

ACKNOWLEDGEMENTS

L. F., acknowledge financial support by the HyPerCell (Hybrid Perovskite Solar Cells) joint Graduate School.

Electronic Supplementary Information for

# Influence of Chloride Substitution on the Rotational Dynamics of Methylammonium in MAPbI$_{3-x}$Cl$_x$ Perovskites

Götz Schuck[1]*, Frederike Lehmann[1,2], Jacques Ollivier[3], Hannu Mutka[3], and Susan Schorr[1,4]

[1] Helmholtz-Zentrum Berlin für Materialien und Energie, Hahn-Meitner-Platz 1, 14109 Berlin, Germany
[2] Universität Potsdam, Institut für Chemie, Karl-Liebknecht-Straße 24-25, 14476 Golm, Germany
[3] Institut Laue-Langevin, 71 Avenue des Martyrs, F-38000 Grenoble, France
[4] Institut für Geologische Wissenschaften, Freie Universität Berlin, Malteserstr. 74, 12249 Berlin, Germany

**Table S1** Frequencies [meV] of the peaks in the momentum integrated INS spectra measured on IN5 (4.8 Å incident wavelength). MAPbI$_3$ (red), MAPbCl$_3$ (green) and MAPbI$_{2.94}$Cl$_{0.06}$ (blue) for various temperatures.

| Temp. | P1 | P2 | P3 | P4 | P5 | P6 | P7 | P8 | P9 | P10 | P11 |
|---|---|---|---|---|---|---|---|---|---|---|---|
| | | | | | Peak label | | | | | | |
| | | | | | MAPbI$_3$ | | | | | | |
| 70 K | 2.273(3) | 3.160(6) | 3.747(7) | 4.90(3) | - | 7.5* | 10.69(7) | 11.92(4) | 15.14(6) | 18.3(3) | 37.8(2) |
| 100 K | 2.285(2) | 3.153(5) | 3.748(7) | 4.97(2) | - | 7.5* | 10.71(6) | 12.08(5) | 14.99(5) | 18.2(2) | 37.4(2) |
| 130 K | 2.289(2) | 3.142(5) | 3.720(8) | 4.97(1) | - | 7.5* | 10.69(6) | 12.25(5) | 14.79(6) | 18.1(2) | 37.0(2) |
| 160 K | 2.281(3) | 3.131(8) | 3.68(1) | 4.94(2) | - | 7.5* | 10.62(6) | 12.38(5) | 14.6(1) | 18.0(2) | 36.7(4) |
| | | | | | MAPbI$_{2.94}$Cl$_{0.06}$ | | | | | | |
| 70 K | 2.271(2) | 3.161(5) | 3.754(7) | 5.01(2) | - | 7.5* | 10.70(7) | 11.99(5) | 15.09(4) | 18.3(2) | 37.7(2) |
| 130 K | 2.282(2) | 3.136(4) | 3.713(7) | 5.01(1) | - | 7.5* | 10.64(6) | 12.23(5) | 14.86(5) | 18.2(1) | 36.9(3) |
| | | | | | MAPbCl$_3$ | | | | | | |
| 70 K | 2.57(2) | 3.61(3) | 4.70(1) | 5.37(2) | 6.42(2) | 7.40(5) | 10.64(2) | 13.66(2) | 15.71(2) | 20.1(1) | 38.5(5) |
| 130 K | 2.64(2) | - | 4.5(1) | 5.12(3) | - | 7.72(4) | 10.59(8) | 13.79(4) | - | 20.1(1) | 38.5(5) |

**Table S2** HWHM [meV] of the peaks in the momentum integrated INS spectra measured on IN5 (4.8 Å incident wavelength). MAPbI$_3$ (red), MAPbCl$_3$ (green) and MAPbI$_{2.94}$Cl$_{0.06}$ (blue) for various temperatures.

| Temp. | P1 | P2 | P3 | P4 | P5 | P6 | P7 | P8 | P9 | P10 | P11 |
|---|---|---|---|---|---|---|---|---|---|---|---|
| | | | | | Peak label | | | | | | |
| | | | | | MAPbI$_3$ | | | | | | |
| 70 K | 0.44(1) | 0.19(1) | 0.46(2) | 0.57(7) | - | 5.4(9) | 0.84(9) | 0.92(6) | 1.1(1) | 0.7(6) | 0.6(5) |
| 100 K | 0.463(9) | 0.18(1) | 0.55(2) | 0.43(4) | - | 7.6(3) | 1.10(7) | 1.10(8) | 1.3(1) | 0.9(4) | 1.1(4) |
| 130 K | 0.496(7) | 0.18(2) | 0.67(2) | 0.43(3) | - | 9.1(1) | 1.43(6) | 1.2(1) | 1.6(1) | 1.0(3) | 1.1(6) |
| 160 K | 0.56(1) | 0.22(3) | 0.84(3) | 0.59(5) | - | 9.61(8) | 1.83(6) | 1.4(2) | 2.0(2) | 1.2(4) | (1) |
| | | | | | MAPbI$_{2.94}$Cl$_{0.06}$ | | | | | | |
| 70 K | 0.481(7) | 0.19(1) | 0.59(2) | 0.41(4) | - | 5.6(4) | 0.99(8) | 1.12(6) | 1.3(1) | 1.0(4) | 1.0(4) |
| 130 K | 0.521(5) | 0.19(1) | 0.78(2) | 0.44(3) | - | 6.0(4) | 1.46(6) | 1.5(1) | 1.7(1) | 1.7(3) | 1.1(5) |
| | | | | | MAPbCl$_3$ | | | | | | |
| 70 K | 0.57(7) | 1.1(2) | 0.50(5) | 0.47(4) | 0.45(7) | 1.58(8) | 1.77(5) | 0.86(4) | 1.04(5) | 1.9(2) | 1(1) |
| 130 K | 0.27(5) | - | 6.3(2) | 0.26(8) | - | 1.3(2) | 1.7(2) | 2.31(7) | - | 2.1(3) | 1.1(7) |

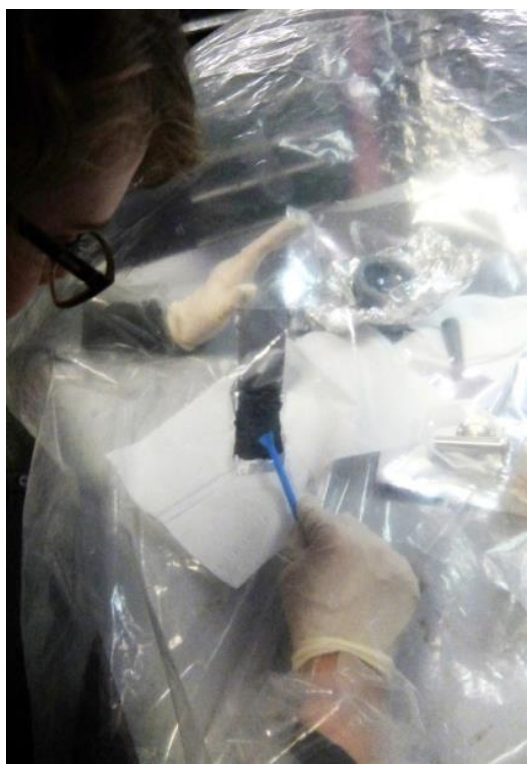

**Fig. S1** All Sample handling was done under Argon atmosphere in order to avoid sample decomposition.

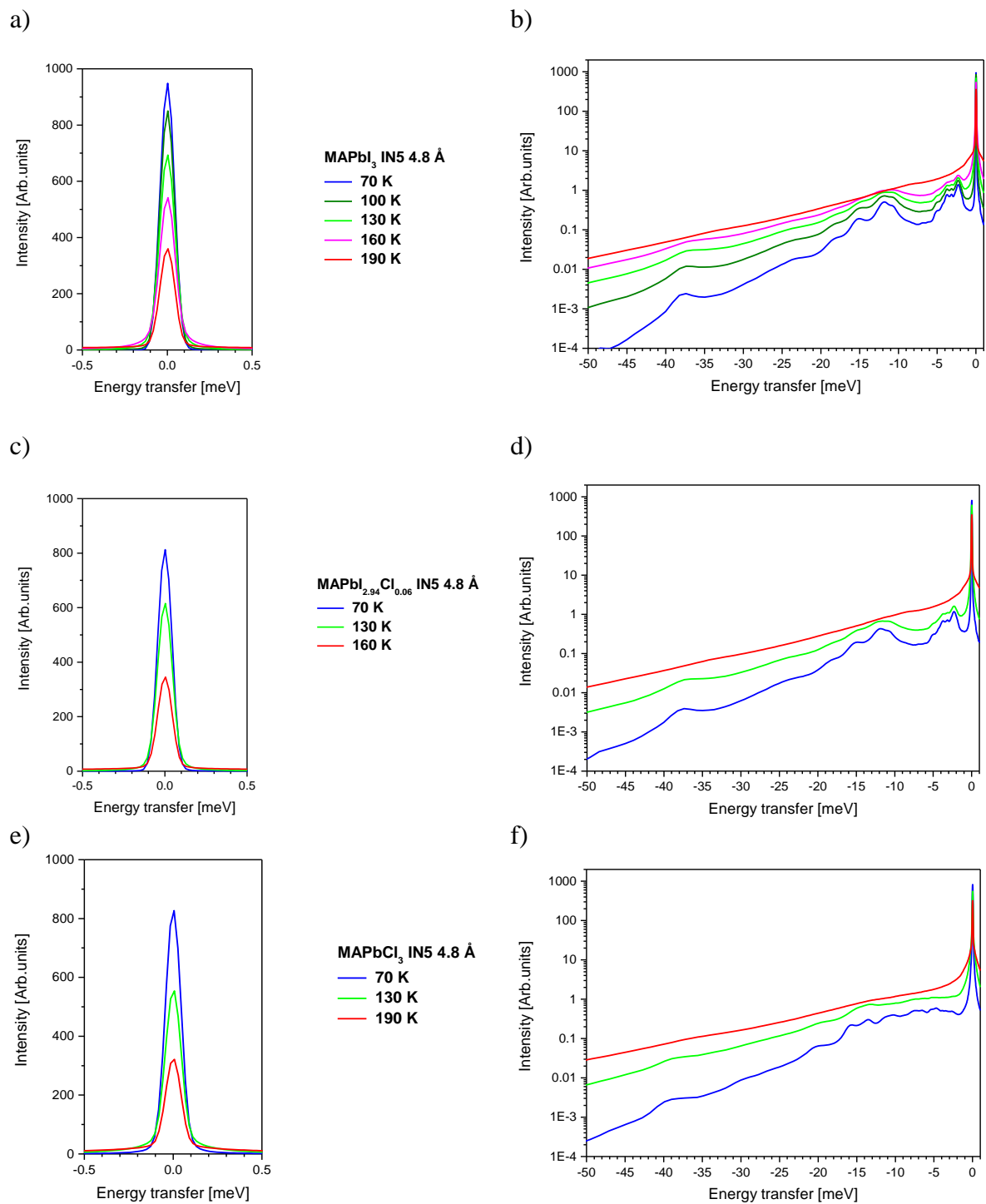

**Fig. S2** Momentum integrated INS spectra measured on IN5 (4.8 Å incident wavelength) of MAPbI$_3$ (a and b), MAPbI$_{2.94}$Cl$_{0.06}$ (c and d) and MAPbCl$_3$ (e and f) at 70 K (blue), 100 K (dark green), 130 K (green), 160 K (magenta), and 190 K (red). a) quasi-elastic scattering as a function of temperature, b) wider energy transfer range, showing low-energy phonons between -35 and -2 meV and internal MA vibrational mode $\nu_6$ (C-N torsion oscillation) at around -37 meV.

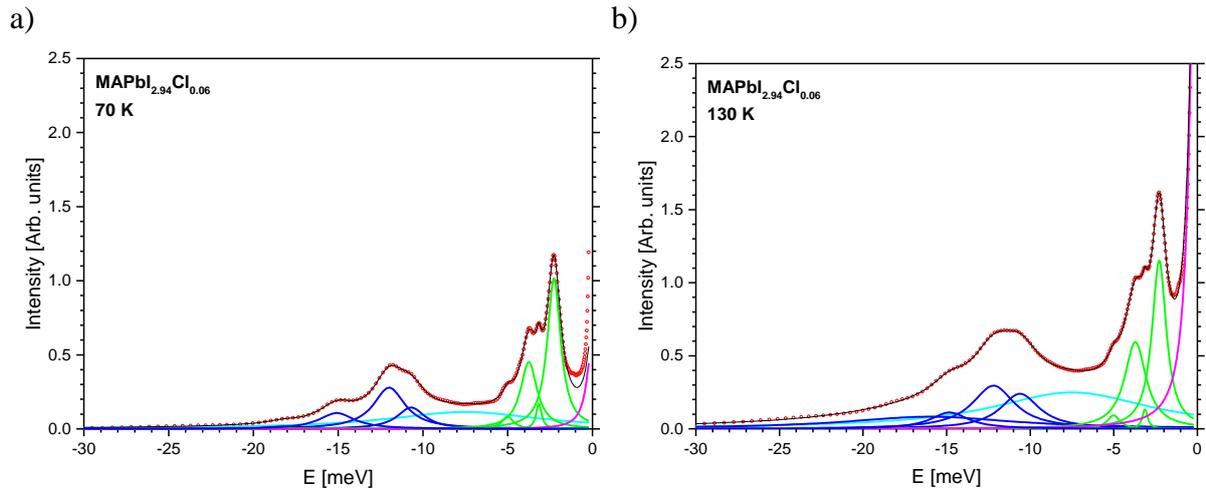

**Fig. S3** Momentum integrated INS spectra measured on IN5 (4.8 Å incident wavelength) of MAPbI$_{2.94}$Cl$_{0.06}$ at 70 K (a), and 130 K (b). Lorentzian functions (decomposition of the vibrational modes according to Pérez-Osorio et. al. [S1]): combination of MA translational and librational lattice vibrations without MA spinning (green), combination of MA translational and librational lattice vibrations with MA spinning (blue), MA lattice vibrations at fixed energy of -7.5 meV (cyan), and modelling the elastic and quasi-elastic neutron scattering at a fixed energy of 0 meV (magenta). Energy range used for the modelling: -30 to -1.5 meV. No background function was used here.

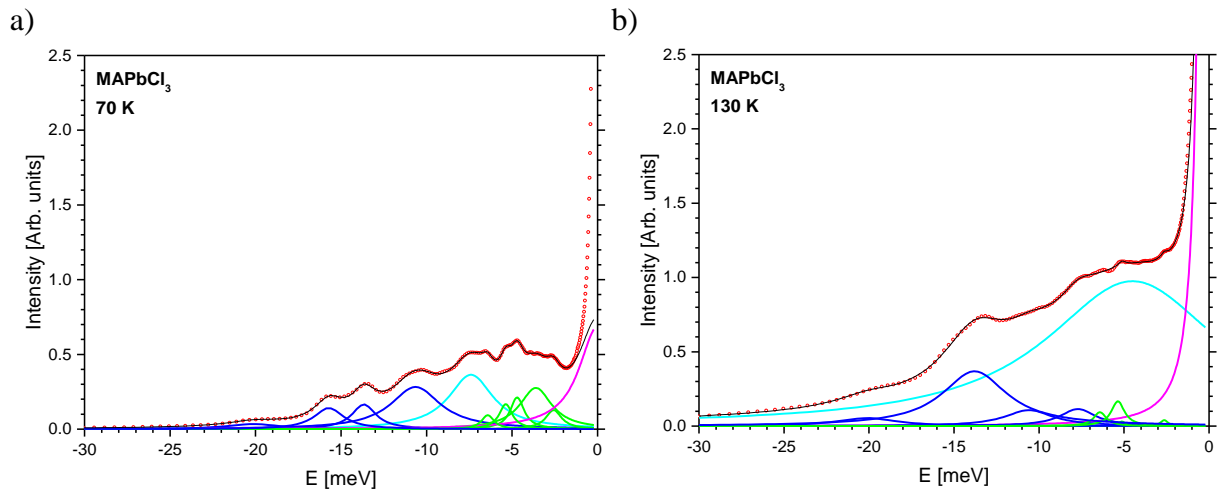

**Fig. S4** Momentum integrated INS spectra measured on IN5 (4.8 Å incident wavelength) of MAPbCl$_3$ at 70 K (a), and 130 K (b). Lorentzian functions (decomposition of the vibrational modes according to Pérez-Osorio et. al. [S1]): combination of MA translational and librational lattice vibrations without MA spinning (green), combination of MA translational and librational lattice vibrations with MA spinning (blue), MA lattice vibrations at fixed energy of -7.5 meV (cyan), and modelling the elastic and quasi-elastic neutron scattering at a fixed energy of 0 meV (magenta). Energy range used for the modelling: -30 to -1.5 meV. No background function was used here.

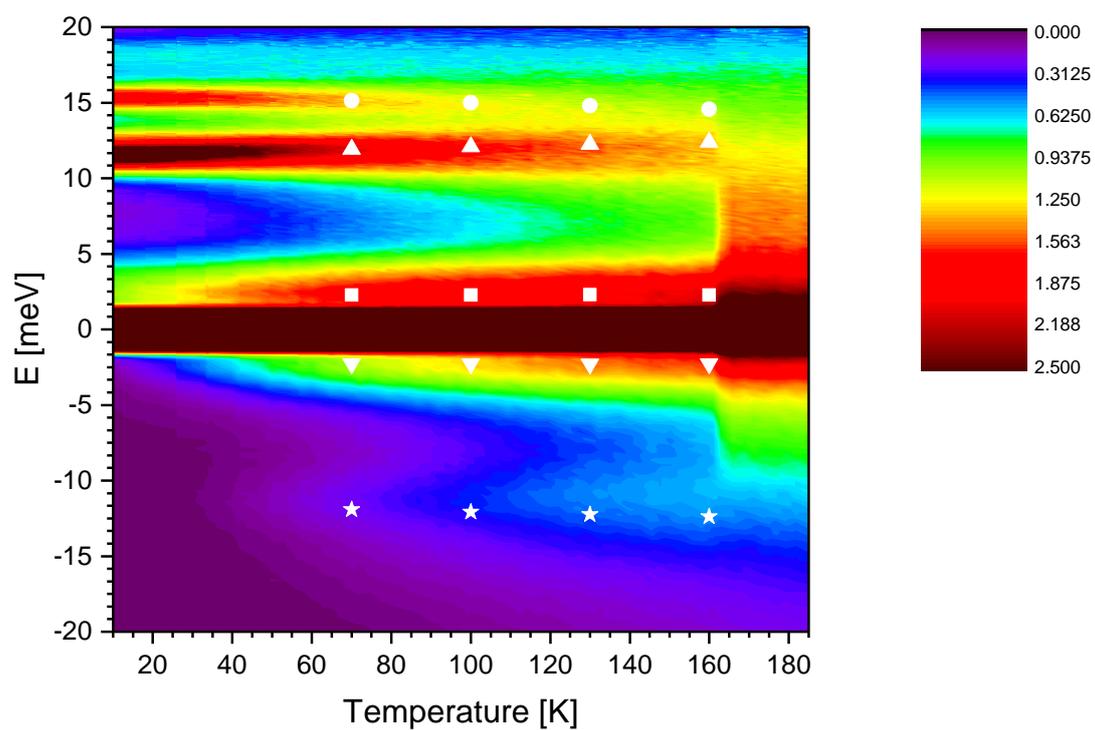

**Fig. S5** Temperature dependent, momentum-transfer averaged spectra of MAPbI$_3$ measured at IN4. For comparison, the frequencies determined in the INS fits (IN5 4.8 Å incident wavelength) were also displayed (white symbols), whereby energy gain and energy loss were displayed accordingly with the same frequency.

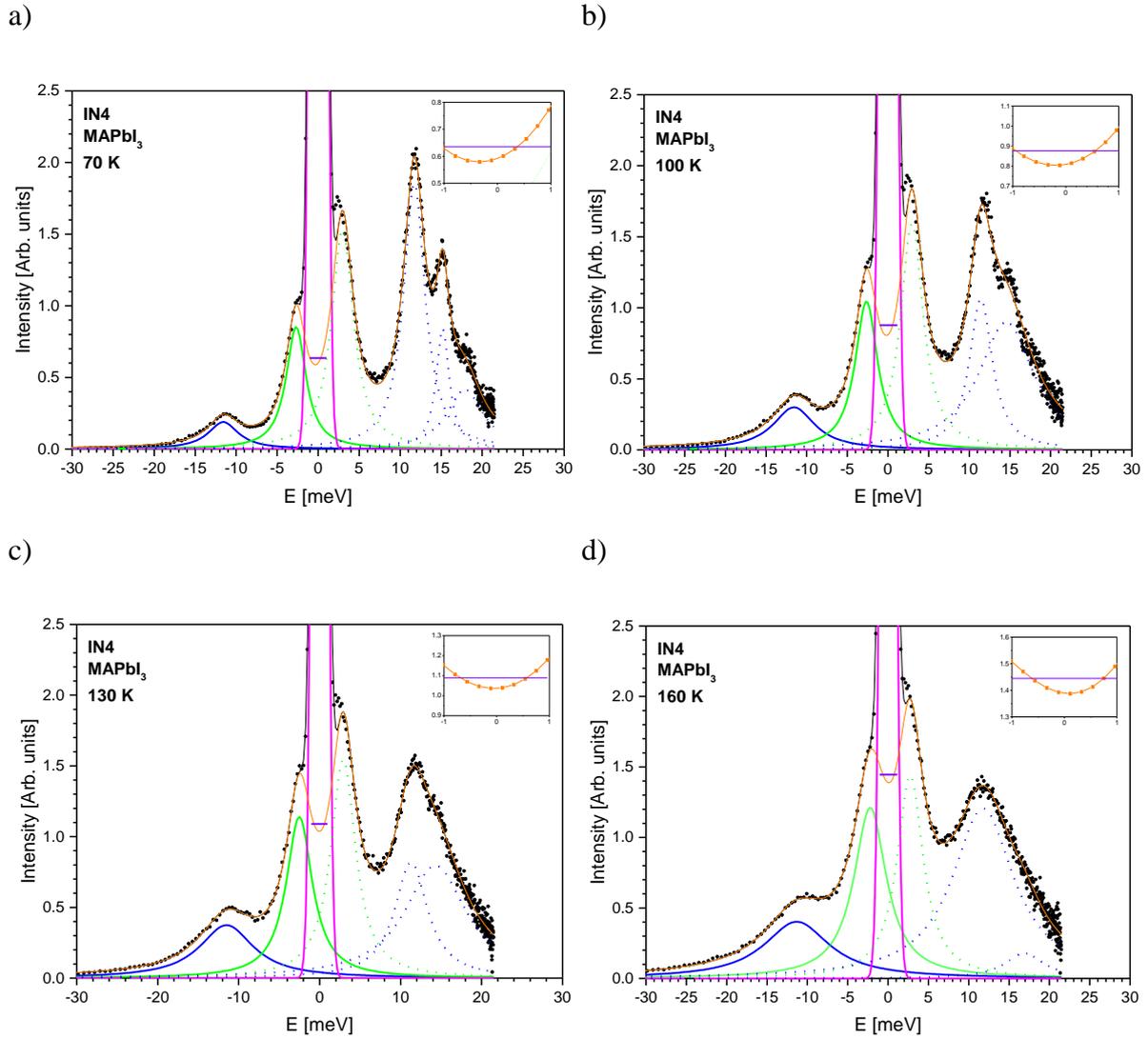

**Fig. S6** Momentum integrated INS spectra measured on IN4 of MAPbI$_3$ at 70 K (a), 100 K (b), 130 K (c), and 160 K (d). Lorentzian functions (neutron energy gain: solid lines; neutron energy loss: doted lines): combination of MA translational and librational lattice vibrations without MA spinning (green), combination of MA translational and librational lattice vibrations with MA spinning (blue), decomposition of the vibrational modes according to Pérez-Osorio et. al. [S1]. Magenta line: modelling the elastic neutron scattering at a fixed energy of 0 meV with a Gaussian function. Energy range used for the modelling: -30 to 21.5 meV. No background function was used here. The sum of the lattice vibrations is shown as an orange colour line, inset shows zoom on the sum of the lattice vibrations and on a straight-line fit of this sum (violet colour); energy range for this fit: -1 meV to 1 meV (the results of this fit as a function of temperature are also shown in figure S7).

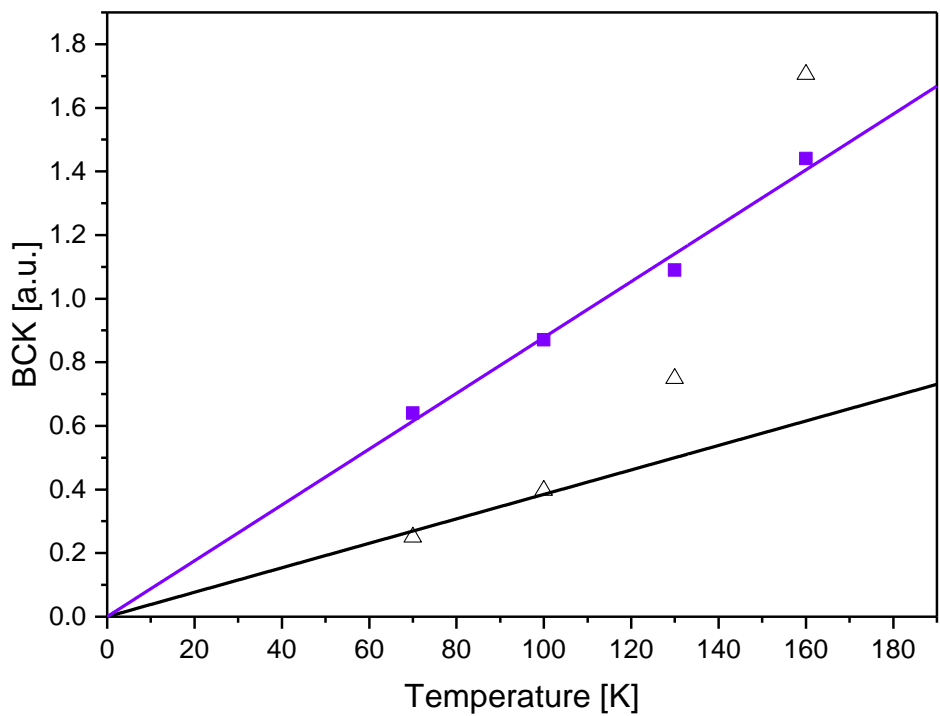

**Fig. S7** Results of a straight-line fit (energy range: -1 meV to 1 meV) of the sum of the lattice vibrations (see figure S6) modelled in momentum integrated INS spectra of MAPbI$_3$ measured on IN4 as a function of temperature (violet square). For comparison the values of momentum averaged flat background values of jumping rotational dynamics C3 QENS model fit of MAPbI$_3$ measured on IN5 is shown (open black triangle). The IN5 flat background values are shown in figure S9a as a function of Q, for various temperatures.

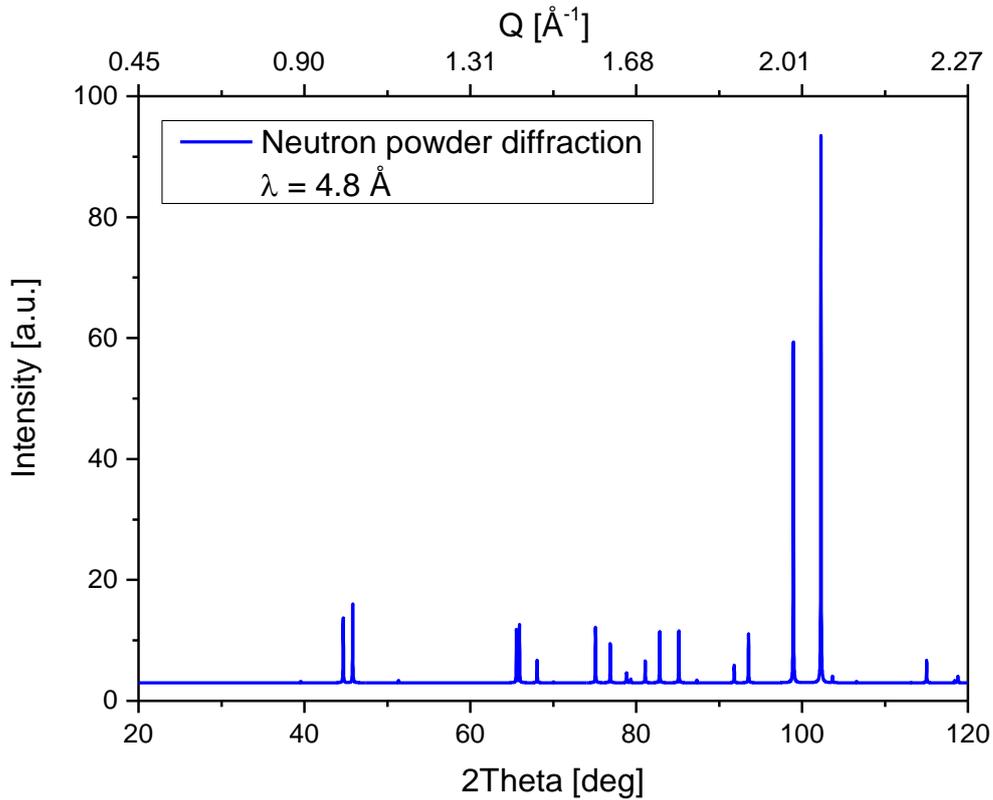

**Fig. S8** MAPbI$_3$ orthorhombic crystal structure,[S2] neutron powder diffraction simulation, $\lambda$ = 4.8 Å.

a)            b)

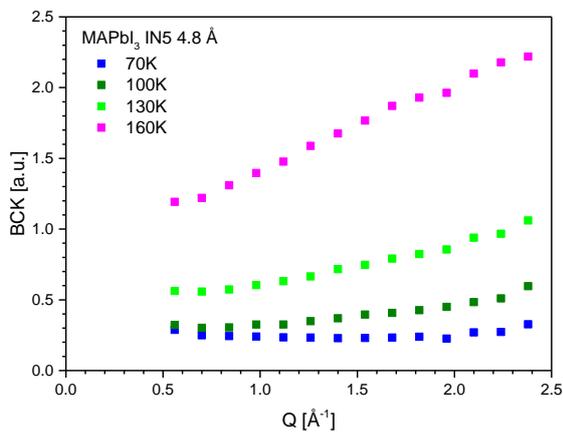 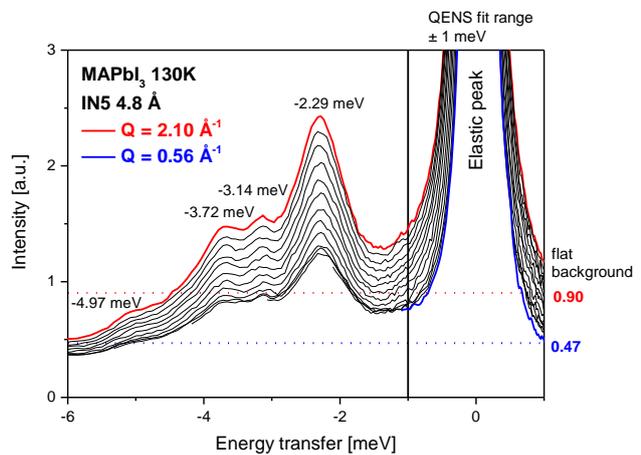

**Fig. S9** a) Constant background as function of Q at different temperatures determined in QENS fits with the C$_3$ model for data measured with MAPbI$_3$ and an incident wavelength of 4.8 Å. b) Q dependent QENS spectra of MAPbI$_3$ at 130 K, the corresponding values for the constant background are shown for Q = 0.56 Å$^{-1}$ and Q = 2.10 Å$^{-1}$. For the QENS fits an energy transfer range of ± 1 meV was used.

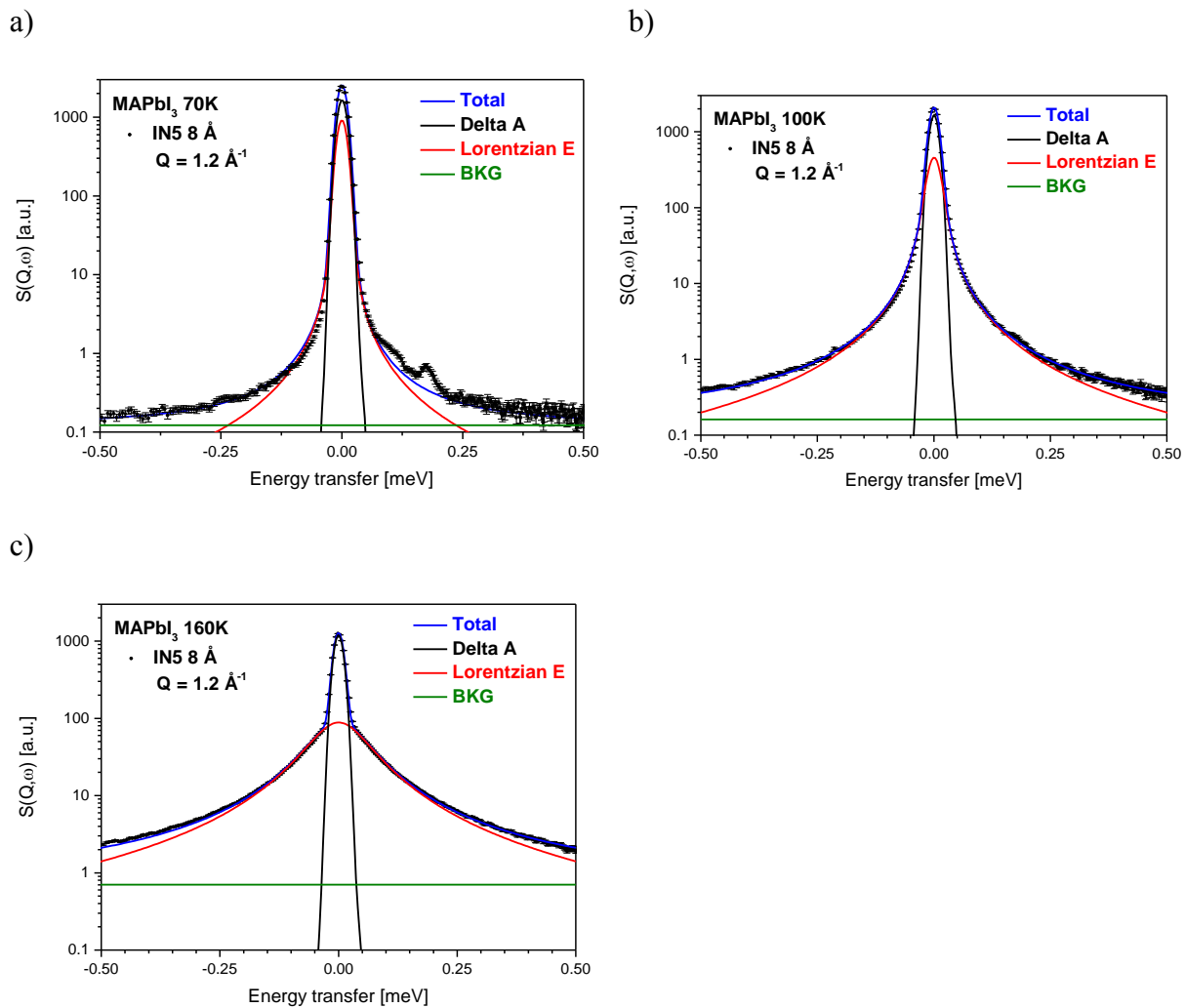

**Fig. S10** Jumping rotational dynamics of MA molecules in MAPbI$_3$. Fit of the C$_3$ model to a spectrum measured at a) 70 K, b) 100 K, and c) 160 K (IN5 8 Å; elastic resolution FWHM = 20 μeV at Q = 1.2 Å$^{-1}$). The C$_3$ fit model includes: delta function (black line), Lorentzian function (red line), constant background (green line), and the sum (blue line) which fits the measured spectra (circles).

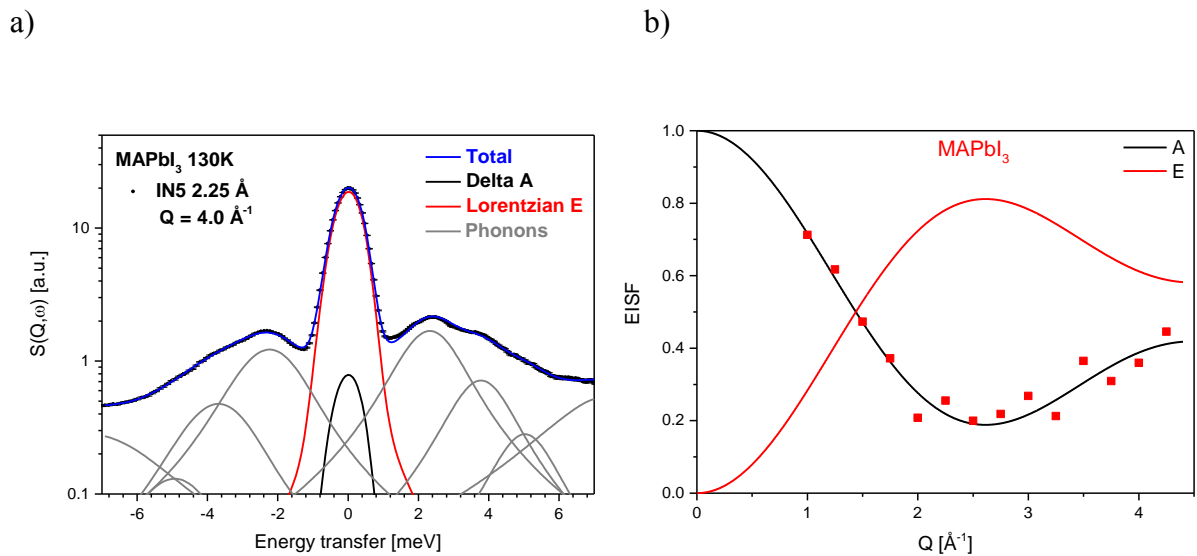

**Fig. S11** Jumping rotational dynamics of MA molecules in MAPbI$_3$. a) Fit of the C$_3$ model to a spectrum measured at 130 K (IN5 2.25 Å; elastic resolution FWHM = 800 μeV at Q = 4.0 Å$^{-1}$). The C$_3$ fit model includes: delta function (black line), Lorentzian function (red line), and the sum (blue line) which fits the measured spectra (circles). Instead of a constant background eight Lorentzian functions (grey line) are utilized to model the phonon induced inelastic neutron scattering. b) The experimental EISF of MAPbI$_3$ as a function of Q (130 K (red squares) for IN5 energy resolution of 2.25 Å). The HWHM Γ$_{C3}$ compiled in Table 1 was used for the fit. The theoretical EISF (A, black line) and QISF (E, red line) behaviour is shown for comparison (the C$_3$ jump model described in section 2.1 was used to calculate the functions).

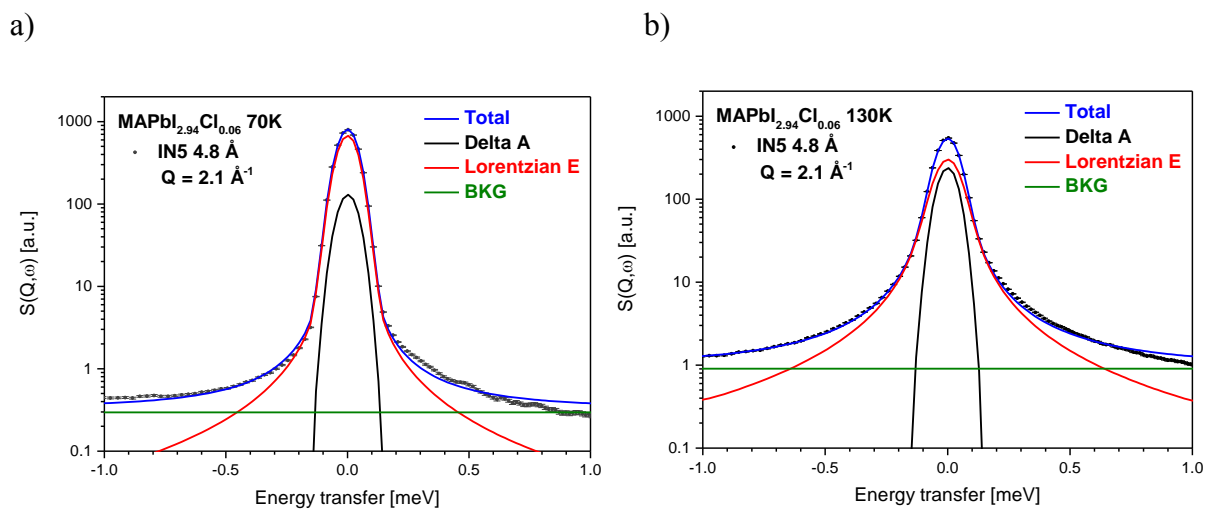

**Fig. S12** Jumping rotational dynamics of MA molecules in MAPbI$_{2.94}$Cl$_{0.06}$. Fit of the C$_3$ model to a spectrum measured at a) 70 K and b) 130 K (IN5 4.8 Å; elastic resolution FWHM = 86 μeV at Q = 2.1 Å$^{-1}$). The C$_3$ fit model includes: delta function (black line), Lorentzian function (red line), constant background (green line), and the sum (blue line) which fits the measured spectra (circles).

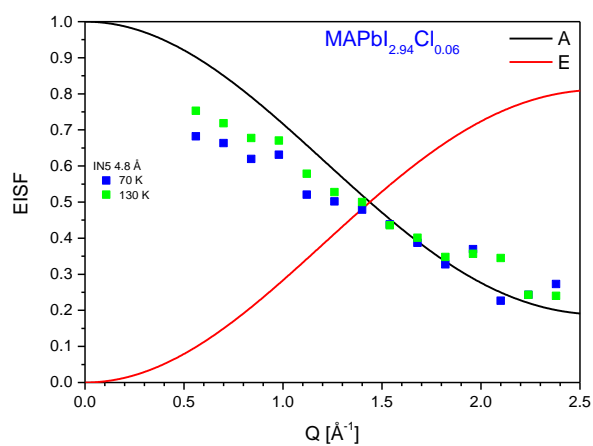

**Fig. S13** The experimental EISF of MAPbI$_{2.94}$Cl$_{0.06}$ as a function of Q. At a temperature of 70 K (blue) and 130 K (green) for IN5 incident wavelength of 4.8 Å (filled symbols). The HWHM $\Gamma_{C3}$ compiled in Table 1 was used for the fit. The theoretical EISF (A, black line) and QISF (E, red line) behaviour is shown for comparison (the C$_3$ jump model described in section 2.1 was used to calculate the functions).

a) 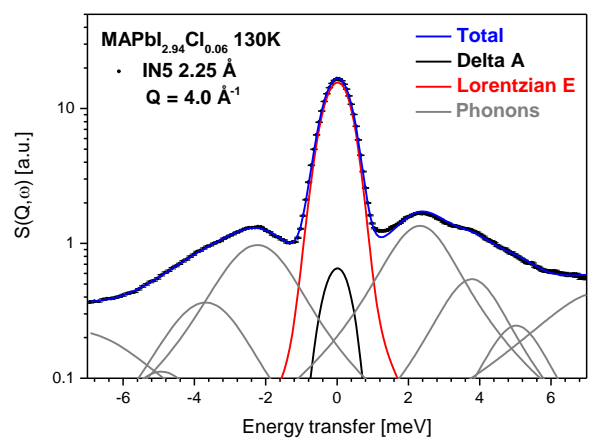 b) 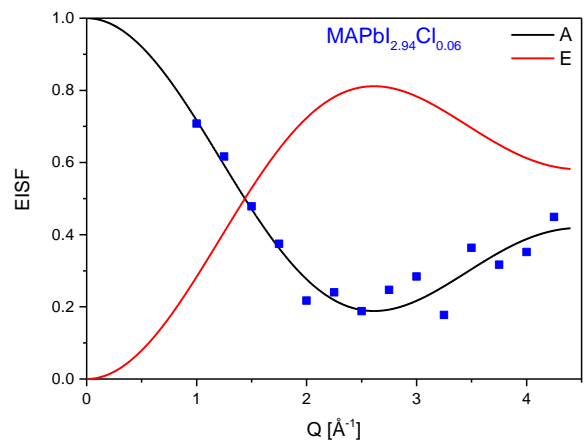

**Fig. S14** Jumping rotational dynamics of MA molecules in of MAPbI$_{2.94}$Cl$_{0.06}$. a) Fit of the C$_3$ model to a spectrum measured at 130 K (IN5 2.25 Å; elastic resolution FWHM = 800 µeV at Q = 4.0 Å$^{-1}$). The C$_3$ fit model includes: delta function (black line), Lorentzian function (red line), and the sum (blue line) which fits the measured spectra (circles). Instead of a constant background eight Lorentzian functions (grey line) are utilized to model the phonon induced inelastic neutron scattering. b) The experimental EISF of of MAPbI$_{2.94}$Cl$_{0.06}$ as a function of Q (130 K (blue squares) for IN5 incident wavelength of 2.25 Å). The HWHM $\Gamma_{C3}$ compiled in Table 1 was used for the fit. The theoretical EISF (A, black line) and QISF (E, red line) behaviour is shown for comparison (the C$_3$ jump model described in section 2.1 was used to calculate the functions).

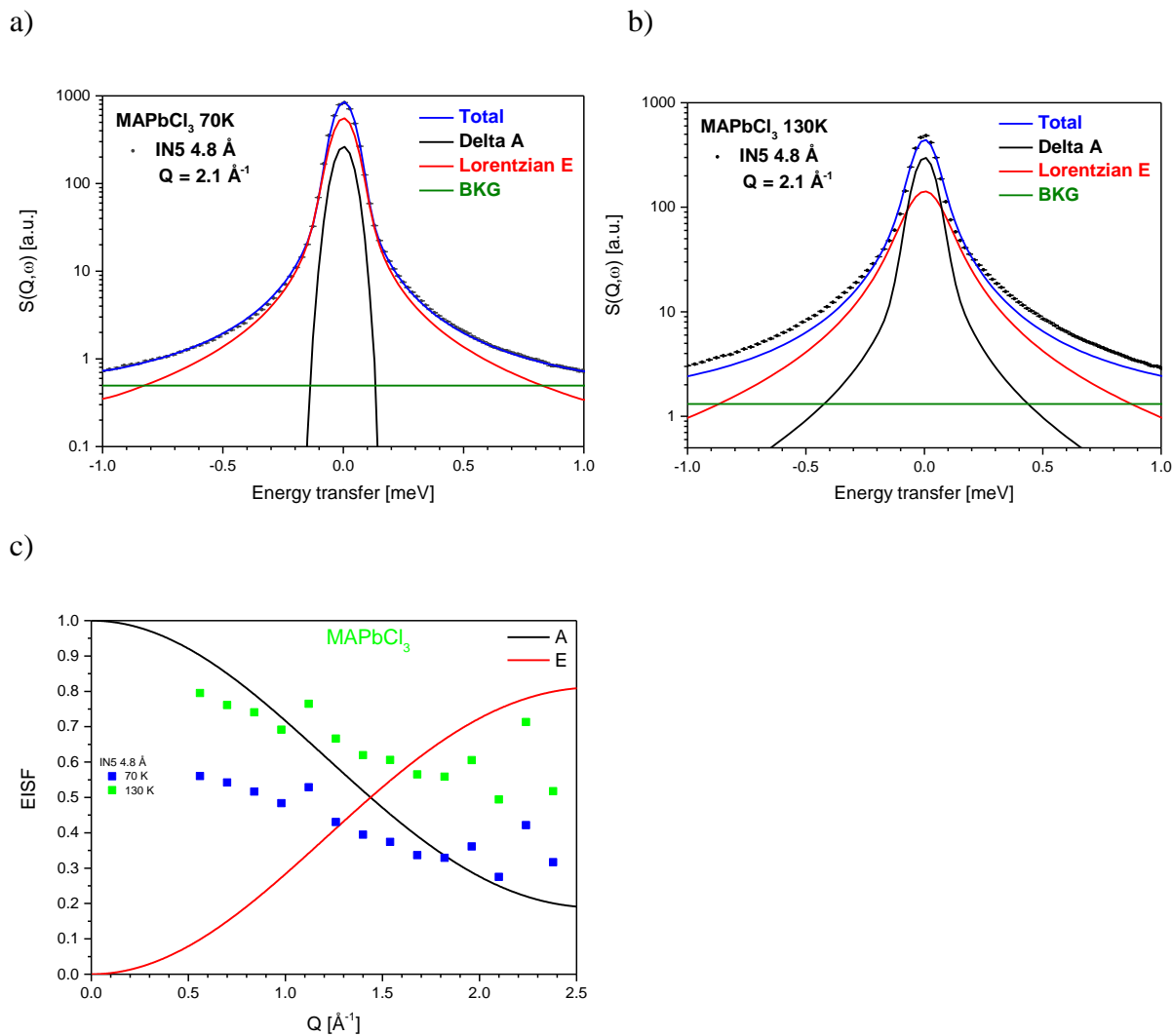

**Fig. S15** Jumping rotational dynamics of MA molecules in MAPbCl$_3$. a) and b): Fit of the C$_3$ model to a spectrum measured at a) 70 K and b) 130 K (IN5 4.8 Å; elastic resolution FWHM = 86 μeV at Q = 2.1 Å$^{-1}$). The C$_3$ fit model includes: delta function (black line), Lorentzian function (red line), constant background (green line), and the sum (blue line) which fits the measured spectra (circles). c): The experimental EISF of MAPbCl$_3$ as a function of Q. At a temperature of 70 K (blue) and 130 K (green) for IN5 incident wavelength of 4.8 Å. The HWHM $\Gamma_{C3}$ compiled in Table 1 was used for the fit. The theoretical EISF (A, black line) and QISF (E, red line) behaviour is shown for comparison (the C$_3$ jump model described in section 2.1 was used to calculate the functions).

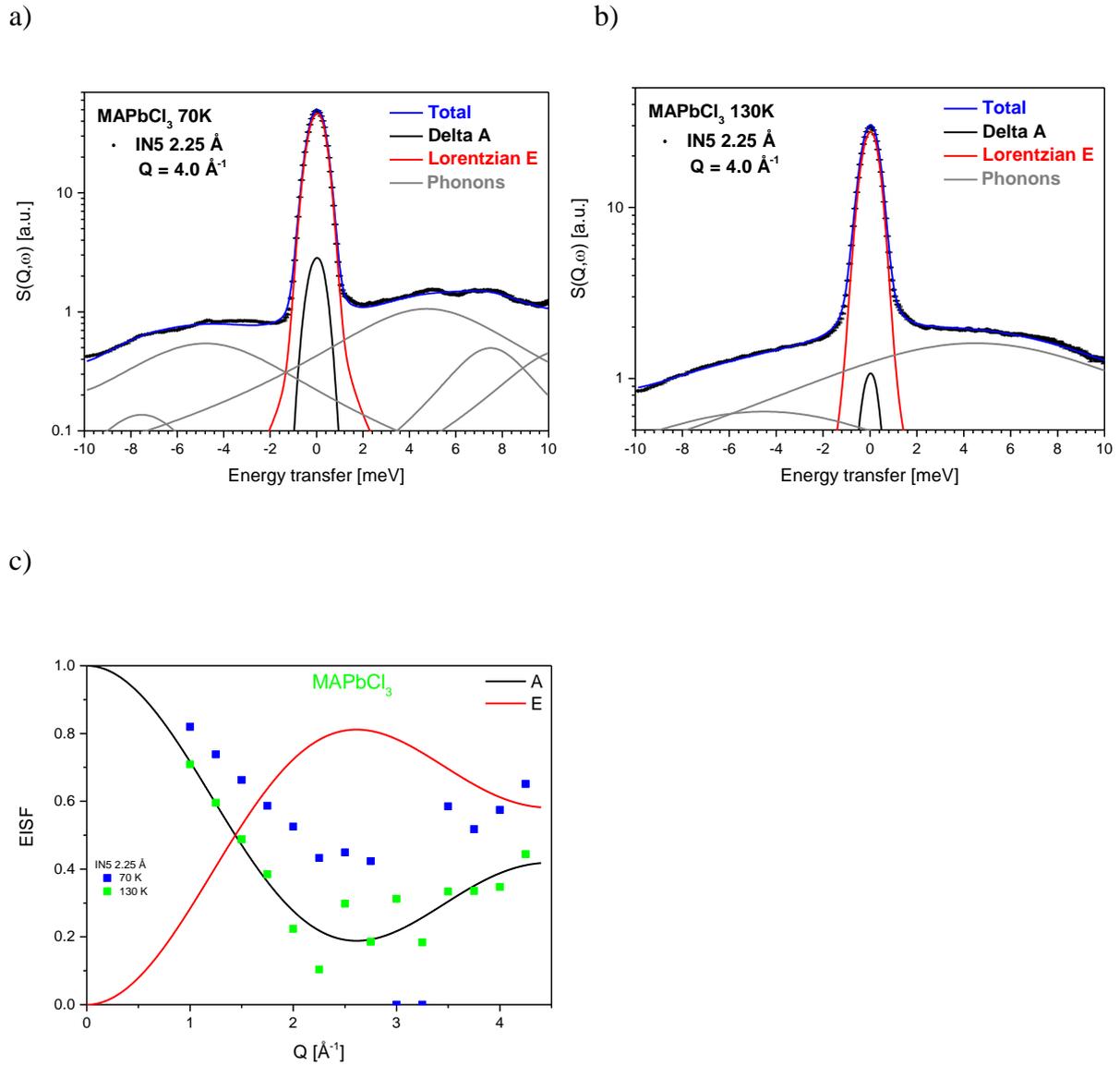

**Fig. S16** Jumping rotational dynamics of MA molecules in MAPbCl$_3$. a) and b): Fit of the C$_3$ model to a spectrum measured at a) 70 K and b) 130 K (IN5 2.25 Å; elastic resolution FWHM = 800 μeV at Q = 4.0 Å$^{-1}$). The C$_3$ fit model includes: delta function (black line), Lorentzian function (red line), and the sum (blue line) which fits the measured spectra (circles). Instead of a constant background five Lorentzian functions for 70 K and two Lorentzian functions for 130 K (grey line) are utilized to model the phonon induced inelastic neutron scattering. c) The experimental EISF of MAPbCl$_3$ as a function of Q (130 K (green squares) and 70 K (blue squares) for IN5 incident wavelength of 2.25 Å). The HWHM $\Gamma_{C3}$ compiled in Table 1 was used for the fit. The theoretical EISF (A, black line) and QISF (E, red line) behaviour is shown for comparison (the C$_3$ jump model described in section 2.1 was used to calculate the functions).

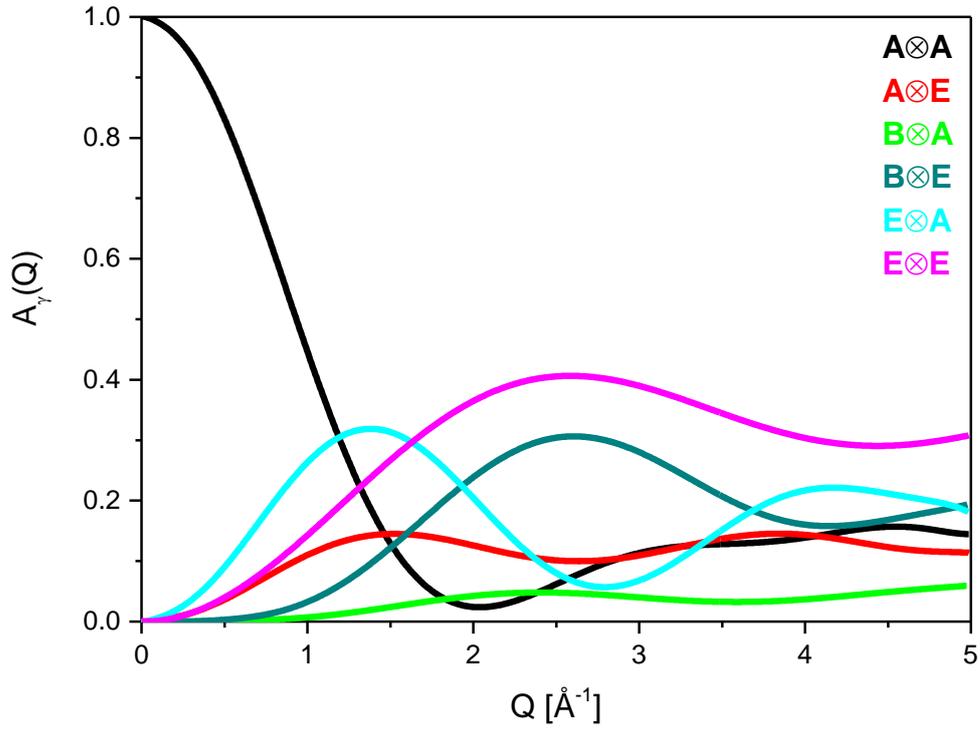

**Fig. S17** Q dependence of the elastic (A⊗A black line) and quasi-elastic incoherent structure factors for the $C_3 \otimes C_4$ model. In the QENS modelling software STRfit a series of polynomial functions was used as a simplified description of this structure factors:

A⊗A: $0.95274+0.56946Q-2.31776Q^2+1.68541Q^3-0.53098Q^4+0.07842Q^5-0.00445Q^6$
A⊗E: $0.00762-0.14889Q+0.65317Q^2-0.5817Q^3+0.21854Q^4-0.03728Q^5+0.00238Q^6$
B⊗A: $0.000748+0.0009098Q-0.0311Q^2+0.06583Q^3-0.03536Q^4+0.00739Q^5-0.000537777Q^6$
B⊗E: $0.12207Q-0.46298Q^2+0.5865Q^3-0.2631Q^4+0.04923Q^5-0.0033Q^6$
E⊗A: $0.0353-0.50353Q+1.9158Q^2-1.70846Q^3+0.63358Q^4-0.10561Q^5+0.00655Q^6$
E⊗E: $0.00159-0.01732Q+0.19498Q^2-0.00743Q^3-0.03823Q^4+0.01065Q^5-0.000828059Q^6$

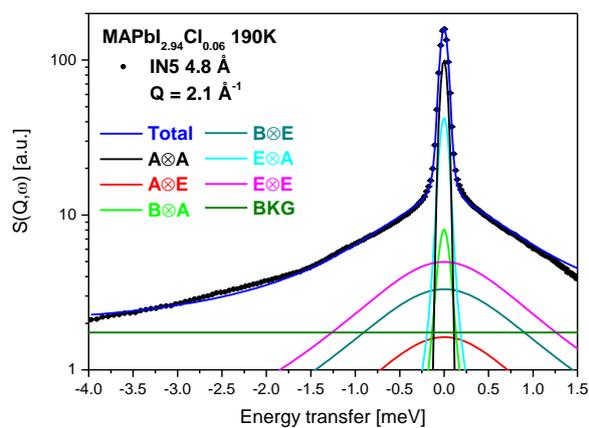 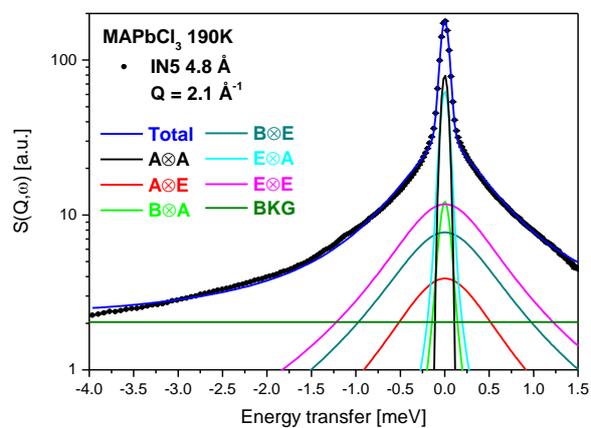

**Fig. S18** Jumping rotational dynamics of MA molecules in a) MAPbI$_{2.94}$Cl$_{0.06}$ and b) MAPbCl$_3$. Fit of the C$_3$⊗C$_4$ model to a spectrum measured at 190 K (IN5 4.8 Å; elastic resolution FWHM = 86 μeV at Q = 2.1 Å$^{-1}$). The C$_3$⊗C$_4$ fit model includes: delta function (black line), five Lorentzian functions, constant background (dark green line), and the sum (blue line) which fits the measured spectra (circles).

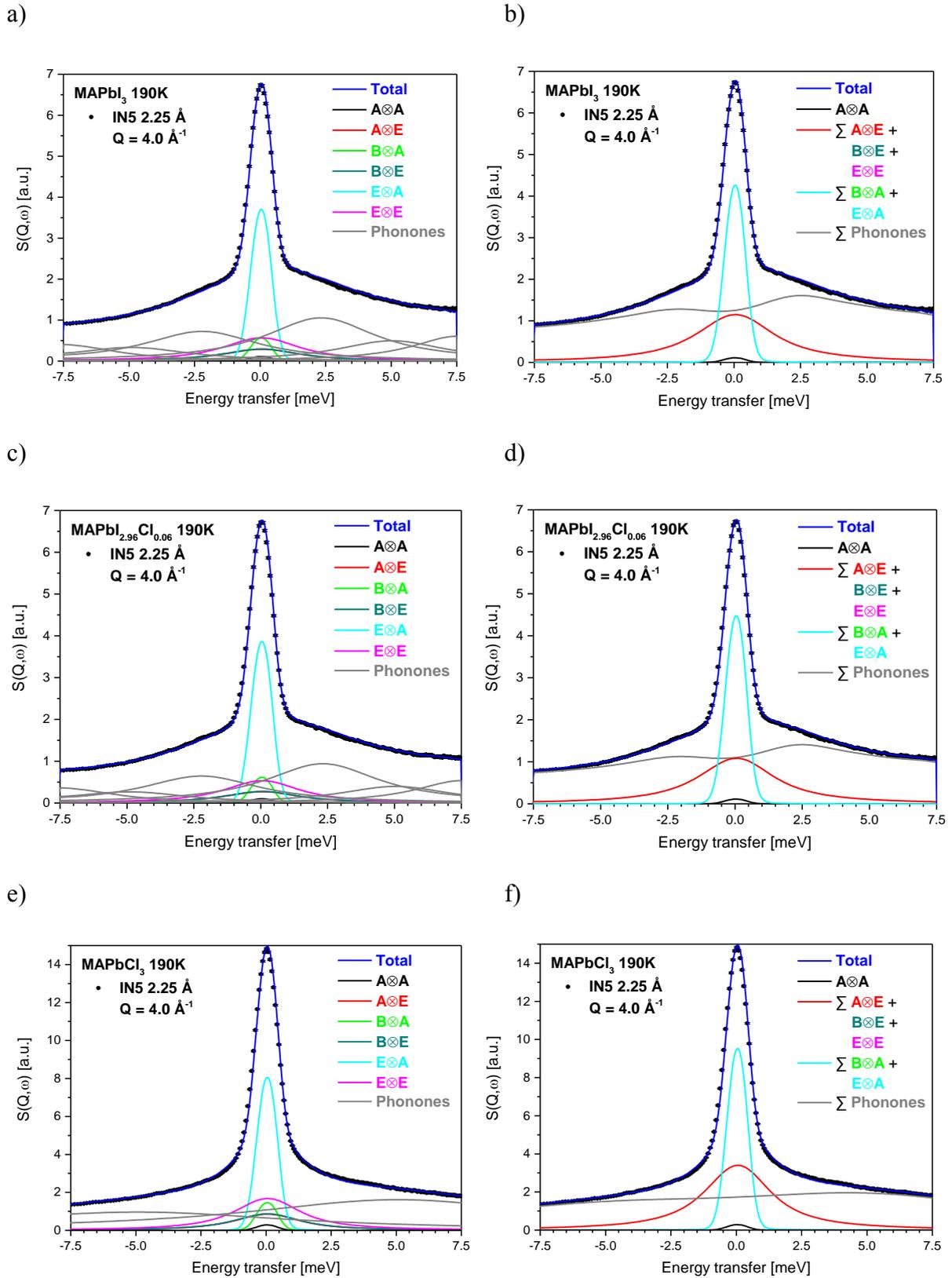

**Fig. S19** Jumping rotational dynamics of MA molecules in a), b) MAPbI$_3$ and c), d) MAPbI$_{2.94}$Cl$_{0.06}$ and e), f) MAPbCl$_3$. Fit of the C$_3\otimes$C$_4$ model to a spectrum measured at 190 K (IN5 2.25 Å; elastic resolution FWHM = 800 μeV at Q = 4.0 Å$^{-1}$). The C$_3\otimes$C$_4$ fit model includes: delta function (black line), five Lorentzian functions, phonon background (gray line), and the sum (blue line) which fits the measured spectra (circles).

a)

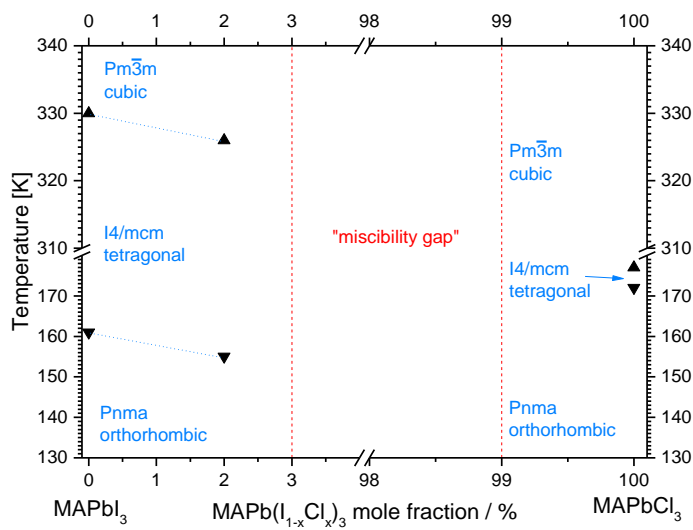

b)

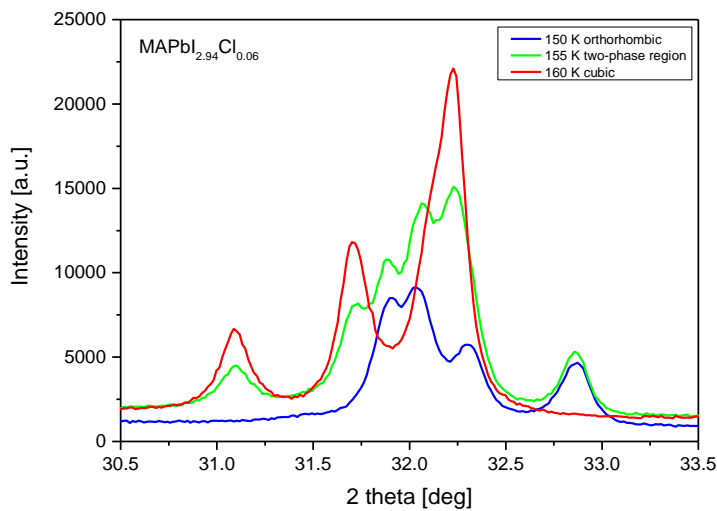

**Figure S20** a) Phase diagram of $MAPbI_{3-x}Cl_x$ solid solution showing the phase transition temperatures over a temperature range from 130 K to 340 K. Recent investigations of X-ray diffraction data were used to establish the solubility limits in $MAPbI_{3-x}Cl_x$.[S3] b) Temperature-dependent synchrotron X-ray diffraction data were used to provide the phase transition temperatures of $MAPbI_{2.94}Cl_{0.06}$. The diffraction data were collected at KMC-2 beamline at the synchrotron source BESSY II (HZB, Berlin, Germany). The data were collected at a photon energy of 8048 eV, corresponding to Cu K$\alpha_1$ radiation.[S4]